\documentclass[12pt]{article}

\usepackage[mathscr]{eucal}
\usepackage{epsfig,amsfonts}
\usepackage{amsmath}
\usepackage{amsthm,amssymb}
\usepackage{bm}
\usepackage{graphicx}
\usepackage{relsize}
\usepackage{hhline}
\usepackage{comment}
\usepackage{bbm}
\usepackage{printlen}
\usepackage{cite}
\usepackage{psfrag}
\usepackage{mathrsfs} 
\usepackage{hyperref}
\usepackage{bm}
\usepackage{array}
\usepackage{tikz}
\usepackage{enumerate}
\usepackage{textcomp}
\usepackage{hyperref}
\usepackage{mathrsfs} 
\usepackage{braket}
\usepackage{float}
\usepackage{enumitem}
\usepackage{subcaption}
\usepackage{ulem}
\numberwithin{equation}{section}
\newcommand{\rulesep}{\unskip\ \vrule\ }
\newcommand{\BR}[2]{u^{[#1]}_{#2}}
\newcommand{\BRv}[2]{v^{[#1]}_{#2}}
\newcommand{\BRvbar}[2]{\bar{v}^{[#1]}_{#2}}
\newcommand{\ri}{{\rm i}}
\newcommand{\re}{{\rm e}}
\newcommand{\id}{\mathbb{I}}

\newcommand{\kk}{{\tt k}}

\makeatletter
\def\@bibdataout@aps{
\immediate\write\@bibdataout{
@CONTROL{
apsrev41Control
\longbibliography@sw{%
    ,author="08",editor="1",pages="1",title="0",year="1"%
    }{%
    ,author="08",editor="1",pages="1",title="",year="1"%
    }%
  }%
}%
\if@filesw \immediate \write \@auxout {\string \citation {apsrev41Control}}\fi 
}
\makeatother
\begin{document}

\begin{titlepage}
$\phantom{I}$
\vspace{2.8cm}

\begin{center}
\begin{LARGE}

{The $D^{(2)}_{3}$ spin chain and its finite-size spectrum}
\end{LARGE}
\vspace{1.3cm}
\begin{large}
{\bf 

Holger Frahm$^1$, Sascha Gehrmann$^1$,\\ Rafael I. Nepomechie$^2$, Ana L. Retore$^3$}

\end{large}

\vspace{1.cm}
$^1$Institut f$\ddot{{\rm u}}$r Theoretische Physik, 
Leibniz Universit$\ddot{{\rm a}}$t Hannover\\
Appelstra\ss e 2, 30167 Hannover, Germany\\\vspace{0.5cm}
$^2$Physics Department, PO Box 248046\\
University of Miami, Coral Gables, FL 33124 USA\\ \vspace{0.5cm}
$^3$Department of Mathematical Sciences, Durham University\\ Durham, DH1 3LE, UK

\vspace{1.0cm}

\end{center}

\begin{center}

\parbox{13cm}{%
\centerline{\bf Abstract} \vspace{.8cm}
Using the analytic Bethe ansatz, we initiate a study of the scaling limit of the quasi-periodic $D^{(2)}_3$ spin chain. Supported by a detailed symmetry analysis, we determine the effective scaling dimensions of a large class of states in the parameter regime $\gamma\in (0,\frac{\pi}{4})$. Besides two compact degrees of freedom, we identify two independent continuous components in the finite-size spectrum. The influence of large twist angles on the latter reveals also the presence of discrete states. This allows for a conjecture on the central charge of the conformal field theory describing the scaling limit of the lattice model.}

\end{center}

\vfill

\end{titlepage}

\section{Introduction}
Low-dimensional critical lattice models whose low-energy behavior is
captured by field theories with a non-compact target space have
attracted increasing attention in recent years.  Such theories may
support continuous spectra of critical exponents, which have been
argued to describe the multifractal scaling of the critical wave
functions at the integer quantum Hall plateau transition and other
disorder-driven quantum phase transitions \cite{Zirn19,EvMi08}. At the
same time, finite lattice realizations of such models with a compact
configuration space may facilitate numerical studies of questions
arising in the context of AdS/CFT dualities \cite{Scho06}.  This is
particularly true if these lattice models are integrable and allow for
a solution of their spectral problem by means of Bethe ansatz methods.

One approach to study two-dimensional disorder problems such as quantum
Hall systems or dirty $d$-wave superconductors is based on
supersymmetric reformulations in terms of spin chains based on supergroup
symmetries \cite{Zirn94,KoMa97,SeFi99,Gruz99}.  A characteristic
feature of these spin chains is that their Hilbert space has a
$Z_2$ staggering, i.e.\ the local degrees of freedom in these
spin chains are different (conjugate) representations of the
underlying algebra on even and odd sites respectively.  Motivated by these findings, integrable deformations of such staggered superspin chains have been constructed
\cite{Gade98,LiFo99,Gade99,EsFS05,FrMa11,FrMa12}.  Beginning
with Ref.~\cite{EsFS05} on an integrable $sl(2|1)$ superspin chain
with alternating three-dimensional quark and antiquark representations as the local
degrees of freedom, finite-size studies of these models have provided evidence for the
presence of continuous components in their spectrum of critical exponents.  The latter manifest themselves through towers with a macroscopic number of energy levels that extrapolate to the same scaling dimension in the thermodynamic limit, but exhibit strong (logarithmic) corrections to scaling that lift their degeneracies in the finite system.
By now the most studied and best understood of the models with such properties
is the staggered six-vertex model related to the antiferromagnetic
Potts model \cite{IkJS08,IkJS12,CaIk13,FrSe14,BaKL21,BKKL21a}, whose
low energy effective theory has been identified to be the
$SL(2,{R})/U(1)$ black hole conformal field theory featuring one compact and one non-compact bosonic degree of freedom \cite{Witten91}.

A common property of the models mentioned above is that they are staggered (either by choosing alternating representations of the underlying algebra, or by considering inhomogeneous shifts of the spectral parameter in the vertex model).  This staggering allows for the construction of a conserved `quasi momentum' operator, which has been crucial in the identification of the scaling limit.  Note however, there also exist translation-invariant models, e.g.\ based on twisted Lie algebras  including several spin chains from the $A^{(2)}_{n}$ series in their regime III \cite{VeJS14,VeJS14a,VeJS16a}, for which signatures for the existence of continuous components of the conformal spectrum have been observed.  Based on finite-size estimates of their central charge and the spectra of elementary excitations, it has been argued that the $A^{(2)}_n$ models allow for two (or more) non-compact degrees of freedom (in addition to the compact ones) in the scaling limit for the higher rank cases with $n>3$.  Due to numerical and analytical difficulties, a reliable description of the conformal spectra in these models has not been possible, though.

Interestingly, the staggered six-vertex model has recently been shown to allow for a mapping to a homogeneous integrable spin chain constructed from the twisted affine $D_2^{(2)}$ Lie algebra \cite{RPJS20}; as a consequence, a quasi momentum can be constructed in one formulation but not in the other.  Such a factorization is not known for the higher-rank models based on $D^{(2)}_{n}$.  Nevertheless, obvious questions to ask are whether these, too, give rise to a series of non-compact CFTs; what is the counting of compact and non-compact degrees of freedom; and finally, what is the operator content of the CFTs describing the low energy spectrum in the scaling limit. 

With this paper, we begin the investigation of these questions for the simplest case beyond the staggered six-vertex model, i.e.\ the $D^{(2)}_3$ spin chain.  Our paper is organized as follows: in Sec. \ref{sec:basics}, after recalling the underlying integrable structures, we construct the transfer matrix of the model subject to generic diagonal twisted boundary conditions, and we identify some of its symmetries.  Generalizing the analytical Bethe ansatz for the periodic case \cite{Resh87}, we obtain the eigenvalues of the transfer matrix and the resulting Hamiltonian with local, i.e.\ nearest-neighbor interactions.  In Sec. \ref{sec:methodology}, combining results from the exact diagonalization of small systems with the numerical solution of the Bethe equations, we identify the root configurations of the low-lying states.  This is then used in Sec. \ref{sec:ground}
to compute the ground state energy density in the thermodynamic limit, and in Sec. \ref{sec:fsize}
to construct the renormalization group trajectories for the ground state and excitations used in the finite-size scaling analysis of the spectrum.  The latter is done separately for the compact and the non-compact parts of the spectrum, both with and without the twist.  The flow of the compact modes under the twist resembles that of two compact bosons with compactification radii depending on the anisotropy.  In addition, we observe the emergence of discrete states from  the continuous parts of the spectrum of conformal weights for sufficiently large twists, similar as in other lattice realizations of the black hole CFT. Based on the analytical dependence of these discrete weights on the twist, we conjecture a result for the central charge of the model in its scaling limit which is consistent with that of two copies of the $SL(2,R)/U(1)$ sigma model.

\section{The \texorpdfstring{$D^{(2)}_3$}{D23} spin chain}\label{sec:basics}

In this section, we define the $D^{(2)}_3$ spin chain, describe its symmetries, and review its Bethe ansatz solution.

\subsection{The R-matrix}

The main building block of the $D^{(2)}_3$ integrable quantum spin chain is the $36 \times 36$ $D^{(2)}_3$ R-matrix obtained in \cite{Jimb86}. We follow here the conventions in Eqs. (A.8-10) of \cite{NePR17}; i.e.
our R-matrix $\mathbb{R}(u)$ is related to the one of Jimbo $R_{J}(x)$ \cite{Jimb86} by the following identifications and rescaling:
\begin{align}
   \mathbb{R}(u)=\re^{-2u-6\eta}R_{J}(x)\,, \qquad  x=\re^u\,, \qquad k=\re^{2\eta} \,.
\end{align}
Moreover, we henceforth set $\eta = \ri \gamma$, 
so that  $\gamma$ is our anisotropy parameter.
This R-matrix satisfies the Yang-Baxter equation
\begin{equation}
   \mathbb{R}_{12}(u-v)\, \mathbb{R}_{13}(u)\, \mathbb{R}_{23}(v) = \mathbb{R}_{23}(v)\,
   \mathbb{R}_{13}(u)\, 
   \mathbb{R}_{12}(u-v)\,,
   \label{YBE}
\end{equation}
and has the two $U(1)$ symmetries
\begin{equation}
    \left[\mathbb{R}(u) \,, (\mathbbm{h}_j \otimes \id + \id \otimes \mathbbm{h}_j) \right] = 0 \,, \qquad j = 1, 2\,, 
    \label{RU1}
\end{equation}
where 
\begin{equation}
    \mathbbm{h}_1 = e^{(1,1)} - e^{(6,6)} \,, \qquad
    \mathbbm{h}_2 = e^{(2,2)} - e^{(5,5)} \,,
    \label{U1gens}
\end{equation}
and $e^{(k,l)}$ denotes the elementary $6 \times 6$ matrix with 0 everywhere except for one 1 at position $(k,l)$; and $\id$ is the identity matrix.

Further important properties of this R-matrix include PT symmetry:
\begin{equation}
    \mathbb{R}_{21}(u):= \mathcal{P}_{12}\, \mathbb{R}_{12}(u)\, \mathcal{P}_{12} = \mathbb{R}_{12}^{t_1 t_2}(u)\,, 
    \label{PT}
\end{equation}
where $\mathcal{P}$ is the permutation matrix, and $t$ denotes transposition; braiding unitarity:
\begin{equation}
    \mathbb{R}_{12}(u)\, \mathbb{R}_{21}(-u)=\xi(u)\, \xi(-u)\, \id^{\otimes 2}\,, \qquad \xi(u)=4\sinh(u+2\ri \gamma)\sinh(u+4\ri \gamma)\,;
    \label{unitarityR}
 \end{equation}   
regularity:
\begin{equation}    
   \mathbb{R}_{12}(0)=\xi(0)\, \mathcal{P}_{12}\,; \label{Reg}
\end{equation}
crossing symmetry:
\begin{equation}
\mathbb{R}_{12}(u) = V_1\, \mathbb{R}_{12}^{t_2}(4\ri \gamma-u)\, V_1 
= V_2^{t_2}\, \mathbb{R}_{12}^{t_1}(4\ri \gamma-u)\, V_2^{t_2}\,, \qquad 
V = \left(\begin{smallmatrix}
     &  &  &   &  &  \re^{-3\ri \gamma}\\
     &  &  &   &  \re^{-\ri \gamma} & \\
     &  &  &   1 &   &   \\
     &  &  1 &  &    & \\
     &  \re^{\ri \gamma} &  &  &  &   \\
     \re^{3\ri \gamma} &  &  &  &  &  
    \end{smallmatrix}\right) \,,
    \label{crossingR}
\end{equation}
with $V^2 = \id$ and 
\begin{equation}
 V_1\, \mathbb{R}_{12}(u)\, V_1= V_2\, \mathbb{R}_{21}(u)\, V_2 \,;
 \label{Vprop}
\end{equation}
quasi-periodicity:
\begin{equation}
    \mathbb{R}_{12}(u+i\pi) = U_1\, \mathbb{R}_{12}(u)\, U_1 = U_2\, \mathbb{R}_{12}(u)\, U_2\,, \qquad U = \left(\begin{smallmatrix}
    1 & & & & &\\
    & 1 & & & &\\
    & & 0 & 1 & &\\
    & & 1 & 0 & &\\
    & & & & & 1 & \\
    & & & & & & 1
    \end{smallmatrix}\right) \,,
\label{quasiperiodicity}
\end{equation}
with $U^2 = \id$;
and the two $Z_2$ symmetries:
\begin{align}
    \mathbb{R}_{12}(u) &= U_1\,  U_2\, \mathbb{R}_{12}(u)\, U_1 U_2\,, 
    \label{Usymmetry}\\\
    \mathbb{R}_{12}(u) &= W_1(u)\, W_2(0)\,  \mathbb{R}_{12}(u)\, W_1(u)\, W_2(0)\,, \qquad W(u)= \left(\begin{smallmatrix}
     &  &  &   & \re^{-u} & \\
     &  &  &   &    & -\re^{-u}\\
     &  & 1 &   &    & \\
     &  &  & -1 &    & \\
    \re^{u}&  &  &  &  &   \\
     & -\re^{u} &  &  &  &  
    \end{smallmatrix}\right) \,.
\label{RWsymmetry}
\end{align}
with $W(u)^2 = \id$.

\subsection{The transfer matrix and its symmetries}

We consider a closed homogeneous spin chain of length $L$
with diagonally-twisted boundary conditions.
The corresponding transfer matrix $\mathbbm{t}(u)$ is therefore given by \cite{Vega84, Skly87}
\begin{align}
    \mathbbm{t}(u):=\mathrm{tr}_{0}\, \mathbb{K}_0\, \mathbb{T}_0(u)\,,
    \label{twistedtransf}
\end{align}
where $\mathbb{T}_{0}(u)$ is the monodromy matrix
\begin{equation}
    \mathbb{T}_{0}(u):=\mathbb{R}_{0L}(u)\dots \mathbb{R}_{01}(u)\,.
    \label{monodromy}
\end{equation}
Moreover, the diagonal twist matrix $\mathbb{K}$ is given by
\begin{equation}
    \mathbb{K} = \re^{\sum_{j=1}^2 \ri \phi_j\, \mathbbm{h}_j}
    = \text{diag}\left(\re^{\ri \phi_1}\,, \re^{\ri \phi_2}\,,1\,,1\,, \re^{-\ri \phi_2}\,, \re^{-\ri \phi_1}\right)\,,
    \label{twistmatrix}
\end{equation}
where $\phi_1$ and $\phi_2$ are twist angles, which we restrict to real $\phi_{1,2}$.
Note that the local Hilbert space at each site
has dimension 6. Furthermore,
\begin{equation}
    \left[ \mathbb{R}(u) \,, \mathbb{K} \otimes \mathbb{K} \right] =  \left[ \mathbb{R}(u) \,,  \re^{\sum_{j=1}^2 \ri \phi_j\, (\mathbbm{h}_j \otimes \id + \id \otimes \mathbbm{h}_j)}\right] = 0 \,,
    \label{RKK}
\end{equation}
where the last equality follows from \eqref{RU1}.
The transfer matrix \eqref{twistedtransf} has the commutativity property
\begin{equation}
    \left[ \mathbbm{t}(u) \,, \mathbbm{t}(v) \right] = 0 \,,\label{TT}
\end{equation}
which is the hallmark of quantum integrability,
as a consequence of \eqref{YBE}, \eqref{RKK}. 

The $U(1)$ symmetries of the R-matrix \eqref{RU1}
are inherited by the transfer matrix
\begin{equation}
    \left[ \mathbbm{t}(u)\,, \mathbbm{h}_j^{(L)} \right] = 0 \,, \qquad j = 1, 2 \,,
    \label{U1transf}
\end{equation}
where\footnote{We will henceforth abbreviate $\mathbbm{h}_j^{(L)}$ as $\mathbbm{h}_j$ when the meaning is clear from the context.} 
\begin{equation}
   \mathbbm{h}_j^{(L)} = \sum_{i=1}^L \left( \mathbbm{h}_j \right)_i \,,
\end{equation}
and $\left( \mathbbm{h}_j \right)_i$ denotes the generator $\mathbbm{h}_j$ \eqref{U1gens} at site $i$, that is
\begin{equation}
    \left( \mathbbm{h}_j \right)_i = \id \otimes  \cdots \otimes \id \otimes \underbrace{\mathbbm{h}_j}_{i} \otimes \id \otimes \cdots \otimes\id \,.
\end{equation}

Further properties inherited from the R-matrix
by the transfer matrix include crossing symmetry:
\begin{equation}
     \mathbbm{t}^t(u; \{\phi_j\}) = \mathbbm{t}(4\ri \gamma-u; \{-\phi_j\}) \,,
     \label{crossing}
\end{equation}
periodicity:
\begin{equation}
     \mathbbm{t}(u+i\pi) = \mathbbm{t}(u) \,,
     \label{periodicity}
\end{equation}
and $Z_2$ symmetry:
\begin{equation}
    \mathbbm{t}(u) = U^{\otimes L}\, \mathbbm{t}(u)\, U^{\otimes L} \,,
    \label{Z2symmetry}
\end{equation}
see Eqs. \eqref{crossingR}, \eqref{quasiperiodicity}, \eqref{Usymmetry},   
respectively. The $Z_2$ symmetry \eqref{Z2symmetry} is a 
generalization of the $Z_2$ symmetry found for $D^{(2)}_2$ \cite{IkJS08,RPJS20, NeRe21a}.\footnote{In \cite{NeRe21a}, the $Z_2$ symmetry of the $D^{(2)}_2$ closed-chain transfer matrix is expressed (3.32) in terms of an operator constructed from a matrix $C$ (2.10), in a gauge that is specified by a matrix $B$ (2.9), with
\begin{equation*}
    B\, C\, B = \left(\begin{smallmatrix}
     1 & & & \\
     & 0 & 1 & \\
     & 1 & 0 & \\
     & & & & 1 
    \end{smallmatrix}\right) \,,
\end{equation*}
which is evidently a reduction of the matrix $U$ in \eqref{quasiperiodicity}.} For generic values of the twists $\phi_1, \phi_2$, we also have 
\begin{align}
    W(0)^{\otimes L}\, \mathbbm{t}(u; \phi_1, \phi_2)\, W(0)^{\otimes L}=\mathbbm{t}(u; -\phi_2, -\phi_1) \,.
    \label{W_Sym}
\end{align}
Hence, if the twist angles satisfy $\phi_1 + \phi_2 =0$ (but not for generic values), then the transfer matrix \eqref{twistedtransf} also has the $Z_2$ symmetry 
\begin{align}
    W(0)^{\otimes L}\, \mathbbm{t}(u; \phi, -\phi)\, W(0)^{\otimes L}=\mathbbm{t}(u; \phi, -\phi) \,.
    \label{W_Sym2}
\end{align}

Finally, we note that the transfer matrix has the CPT-like symmetry
\begin{equation}
 V^{\otimes L}\,   \Pi\, \mathbbm{t}^t(u; \{-\phi_j\})\, \Pi\, V^{\otimes L} = \mathbbm{t}(u; \{\phi_j\}) \,,
 \label{CPT}
\end{equation}
where $\Pi$ is the parity operator
\begin{equation}
    \Pi = \prod_{i=1}^{\lfloor\frac{L}{2}\rfloor} \mathcal{P}_{i,L+1-i} \,,
    \label{parityop}
\end{equation}
which on any local operator $X_i$ at site $i$ acts as $\Pi\, X_i\, \Pi = X_{L+1-i}$, see e.g. \cite{DoNe98}; in Eq. \eqref{parityop} $\lfloor x \rfloor$ denotes the floor of $x$.
Proofs of the symmetries \eqref{crossing}, \eqref{W_Sym} and \eqref{CPT} are sketched in Appendix \ref{sec:proofs}.

\subsection{Bethe ansatz}

Let $|\Lambda\rangle$ denote a simultaneous (normalized) eigenstate of the transfer matrix
$\mathbbm{t}(u)$ \eqref{twistedtransf}
and of the $U(1)$ generators $\mathbbm{h}_j$
\begin{align}
    \mathbbm{t}(u)\, |\Lambda\rangle & = \Lambda(u)\,  |\Lambda\rangle \,, \label{simult1} \\
   \mathbbm{h}_j\, |\Lambda\rangle & = h_j\,  |\Lambda\rangle \,, \qquad j = 1, 2\,.
   \label{simult2}
\end{align}
The eigenvalues $\Lambda(u)$ are given by
\begin{align}
\Lambda(u) =& \left(4\sinh(u-2\ri \gamma)\sinh(u-4\ri \gamma)\right)^{L} \re^{\ri \phi_1}\, A(u) \nonumber \\
 &+ \left(4 \sinh(u-4\ri \gamma) \sinh u\right)^L \left[ \re^{\ri \phi_2}\, B_1(u) + B_2(u) + B_3(u) + \re^{-\ri \phi_2}\, B_4(u) \right] \nonumber\\
&+ \left(4\sinh(u-2\ri \gamma)\sinh u\right)^{L} \re^{-\ri \phi_1}\, C(u) \,,
\label{TQtwist}
\end{align}
with 
\begin{equation}
    \label{AB1B2}
\begin{aligned}
    A(u) &= \prod_{j=1}^{m_1} \frac{\sinh(u-u^{[1]}_j+\ri \gamma)}{\sinh(u-u^{[1]}_j-\ri \gamma)}\,,\\
    B_1(u) &= \prod_{j=1}^{m_1}
      \frac{\sinh(u-u^{[1]}_j-3\ri \gamma)}{\sinh(u-u^{[1]}_j-\ri \gamma)}\,
      \prod_{j=1}^{m_2}\frac{\sinh(u-u^{[2]}_j)}{\sinh(u-u^{[2]}_j-2\ri \gamma)}\,, \\
          B_2(u) &      = \prod_{j=1}^{m_2} \frac{2\cosh\left(\frac{1}{2}\left(u-u^{[2]}_j\right)\right)\,\sinh\left(\frac{1}{2}\left(u-u^{[2]}_j-4\ri \gamma\right)\right)}{\sinh(u-u^{[2]}_j-2\ri \gamma)}\,,
\end{aligned}
\end{equation}
and
\begin{equation}
    C(u)=\bar{A}(4\ri \gamma-u)\,, \qquad
    B_3(u)=\bar{B_2}(4\ri \gamma-u)\,, \qquad
    B_4(u)=\bar{B_1}(4\ri \gamma-u)\,, 
\label{CB3B4}
\end{equation}
where the barred quantities are obtained by negating all the Bethe-roots (i.e. $u^{[l]}_j \mapsto -u^{[l]}_j$), that is,
\begin{equation}
    \begin{aligned}
    \bar{A}(u) &= \prod_{j=1}^{m_1} \frac{\sinh(u+u^{[1]}_j+\ri \gamma)}{\sinh(u+u^{[1]}_j-\ri \gamma)}\,,\\
    \bar{B}_1(u) &= \prod_{j=1}^{m_1}
      \frac{\sinh(u+u^{[1]}_j-3\ri \gamma)}{\sinh(u+u^{[1]}_j-\ri \gamma)}\,
      \prod_{j=1}^{m_2}\frac{\sinh(u+u^{[2]}_j)}{\sinh(u+u^{[2]}_j-2\ri \gamma)}\,, \\
          \bar{B}_2(u) &      = \prod_{j=1}^{m_2} \frac{2\cosh\left(\frac{1}{2}\left(u+u^{[2]}_j\right)\right)\,\sinh\left(\frac{1}{2}\left(u+u^{[2]}_j-4\ri \gamma\right)\right)}{\sinh(u+u^{[2]}_j-2\ri \gamma)}\,.
    \end{aligned}
\end{equation}
If $u$ and $\eta=\ri \gamma$ are real, then the bar has the interpretation of complex conjugation, and the periodic transfer-matrix eigenvalue has the crossing symmetry
\begin{equation}
    \bar{\Lambda}(u) = \Lambda(4\ri \gamma-u) \,,
\end{equation}
similarly to the XXZ model \cite{Resh83}.
The result for the periodic case ($\phi_1=\phi_2=0$) was obtained by Reshetikhin  via the analytical Bethe ansatz in \cite{Resh87}. The generalization to the twisted case is presented in Appendix \ref{sec:twistBA}.
Note that we have redefined the Bethe-roots from \cite{Resh87} as
\begin{align}
    u^{[1]}_j=2\ri x_j \,, \qquad u^{[2]}_j=2\ri y_j \,.
\end{align}
We also note that the eigenvalues \eqref{TQtwist} have the periodicity $\Lambda(u+i\pi) = \Lambda(u)$, consistent with \eqref{periodicity}.

An eigenvalue $\Lambda(u)$ of the transfer matrix is a Fourier polynomial and hence an analytic function. By requiring that the residues of \eqref{TQtwist} at the apparent poles
\begin{equation}
  u= \BR{1}{j}+\ri \gamma\,, \qquad
  u= \BR{2}{j}+2\ri \gamma
\end{equation}
vanish, one obtains the Bethe equations (BE) 
\begin{equation}
    \begin{aligned}\label{Twisted_BAE}
    &\left( \frac{\sinh\left(\BR{1}{j}-\ri \gamma\right)}{\sinh\left(\BR{1}{j}+\ri \gamma\right)} \right)^L
    = \re^{\ri(\phi_2 - \phi_1)}\prod_{k\neq j}^{m_1} \frac{\sinh\left(\BR{1}{j}-\BR{1}{k}-2\ri \gamma\right)}{\sinh\left(\BR{1}{j}-\BR{1}{k}+2\ri \gamma\right)}\,
      \prod_{k=1}^{m_2} \frac{\sinh\left(\BR{1}{j}-\BR{2}{k}+\ri \gamma\right)}{\sinh\left(\BR{1}{j}-\BR{2}{k}-\ri \gamma\right)}\,,  \\
  & \qquad\qquad\qquad\qquad\qquad\qquad\qquad\qquad\qquad\qquad\qquad\qquad\qquad\qquad j=1,\dots,m_1\,,\\
    &\prod_{k=1}^{m_1} \frac{\sinh\left(\BR{2}{j}-\BR{1}{k}-\ri \gamma\right)}{\sinh\left(\BR{2}{j}-\BR{1}{k}+\ri \gamma\right)}
    = \re^{-\ri \phi_2}\prod_{k\neq j}^{m_2} \frac{\sinh\frac12\left(\BR{2}{j}-\BR{2}{k}-2\ri \gamma\right)}{\sinh\frac12\left(\BR{2}{j}-\BR{2}{k}+2\ri \gamma\right)}\,,
    \quad j=1,\dots,m_2\,,
    \end{aligned}
\end{equation}
see also \cite{Resh87,VeLo91a}.

Note that the BE are invariant under the  transformations $\BR{1}{j}\to \BR{1}{j}+\ri \pi$ and $\BR{2}{j}\to \BR{2}{j}+2\ri \pi$. Hence, one can restrict 
\begin{align}
    -\frac{\pi}{2}<\Im m(\BR{1}{j})\leq\frac{\pi}{2} \,, \\
      -\pi<\Im m(\BR{2}{j})\leq\pi \,.
\end{align}
Notice that we are also allowed to do $\BR{2}{j}\to \BR{2}{j}+\ri \pi $, but only if we shift all the Bethe-roots (i.e. for every $j$) at the same time.

The eigenvalues $h_j$ of the $U(1)$ generators $\mathbbm{h}_j$ (see Eq. \eqref{simult2}) are given by \cite{Resh87}
\begin{equation}
\label{hvals}
\begin{aligned}
    h_1&=L-m_1 \,, \\
    h_2&=m_1-m_2 \,.
\end{aligned}
\end{equation}
As usual, the Bethe ansatz provides solutions for 
\begin{align}
     L\ge m_1\ge m_2\ge0 \,.
     \label{NumBRConstraint}
\end{align}

Although an algebraic Bethe ansatz construction of the eigenstates $| u^{[1]}_1, \ldots, u^{[1]}_{m_1}\,; u^{[2]}_1, \ldots, u^{[2]}_{m_2} \rangle$ of the $D^{(2)}_3$ transfer-matrix \eqref{twistedtransf} has not yet been worked out in detail, we expect that these states can be constructed as follows: the first level of nesting (introducing type-1 Bethe roots $u^{[1]}_1, \ldots, u^{[1]}_{m_1}$) can be accomplished \cite{Mart99} using certain elements of the monodromy matrix \eqref{monodromy}, reducing the problem to $D^{(2)}_2$. The transfer matrix for the latter can be expressed as a product of $A^{(1)}_1$ transfer matrices \cite{NeRe21a}, which can then be diagonalized by the usual algebraic Bethe ansatz (introducing the type-2 Bethe roots $u^{[2]}_1, \ldots, u^{[2]}_{m_2} $). 
We therefore expect that the $Z_2$ symmetry \eqref{Z2symmetry} shifts all type-2 Bethe roots by $\ri \pi$
\begin{equation}
U^{\otimes L}\, | u^{[1]}_1, \ldots, u^{[1]}_{m_1}\,; u^{[2]}_1, \ldots, u^{[2]}_{m_2} \rangle \propto | u^{[1]}_1, \ldots, u^{[1]}_{m_1}\,; u^{[2]}_1 + \ri\pi, \ldots, u^{[2]}_{m_2} + \ri\pi \rangle \,.
\label{shiftconjecture}
\end{equation}
Indeed, the $Z_2$ symmetry in the $D^{(2)}_2$ case shifts all Bethe roots by $ \ri \pi$ (see (3.34) in \cite{NeRe21a}); and, in the $D^{(2)}_3$ case, these Bethe roots correspond to type-2 Bethe roots.

\subsection{Hamiltonian}

Commuting integrals of motion for the $D^{(2)}_3$ spin chain are obtained by expanding the transfer matrix \eqref{twistedtransf} about the regular point $u=0$ (\ref{Reg}).  The leading term is
\begin{align}
    \mathbbm{t}(0) = \left(4\sinh(2\ri \gamma)\sinh(4\ri \gamma)\right)^L\,\re^{\ri\, \mathbb{P}} \,,
\end{align}
where $e^{\ri\, \mathbb{P}}$ is the one-site translation operator of the model with quasi-periodic boundary conditions, whose matrix elements are given by
\begin{align}
    \left[\re^{\ri\, \mathbb{P}}\right]^{b_1,\dots,b_L}_{a_1,\dots,a_L}=\exp{\left\{\ri \phi_1 \left(\delta^{b_1}_1-\delta^{b_1}_6\right)+\ri \phi_2 \left(\delta^{b_1}_2-\delta^{b_1}_5\right)\right\}}\delta^{b_2}_{a_1}\delta^{b_3}_{a_2} \dots \delta^{b_L}_{a_{L-1}}\delta^{b_1}_{a_L}\,.
    \label{onesiteshift}
\end{align}
From (\ref{TQtwist}) we find that its eigenvalues $\re^{\ri P}$ are parameterized by the Bethe roots as
\begin{align}
    \re^{\ri\, P}=\re^{\ri \phi_1}\,\prod^{m_1}_{k=1} \frac{\sinh(\BR{1}{k}-\ri \gamma)}{\sinh(\BR{1}{k}+\ri \gamma)} \, .
\end{align}
Similarly, we define the local Hamiltonian of the $D^{(2)}_3$ spin chain as\footnote{With this sign the Hamiltonian generalizes the $D^{(2)}_2$ spin chain which has been related to the antiferromagnetic Potts model \cite{RPJS20} or respectively the corresponding staggered-six vertex model \cite{NeRe21a,FrGe22} .}
\begin{align}
  \label{eq:Hamil}
    \mathbb{H}= \sinh(2\ri \gamma) \frac{\mathrm{d}}{\mathrm{d}u} \log\left(\mathbbm{t}(u)\right)\Big\vert_{u=0} +L\sinh(2\ri \gamma)\left[\coth(2\ri \gamma)+\coth(4\ri \gamma)
    \right] \id^{\otimes L}\,.
\end{align}
the eigenenergies are
\begin{align}
    E=\sum^{m_1}_{k=1} \epsilon_0(u^{[1]}_k)=-\sum^{m_1}_{k=1}\frac{2\sinh^2(2\ri \gamma)}{\cosh\left(2\BR{1}{k}\right)-\cosh\left(2\ri \gamma\right)}\,.\label{Energy}
\end{align}

The Hamiltonian of course inherits the $U(1)$ and $Z_2$ symmetries of the transfer matrix
\begin{align}
    \left[ \mathbb{H} \,, \mathbbm{h}_j \right] &= 0\,, \qquad j=1\,,2 \,, \label{U1Ham} \\
     \left[ \mathbb{H}\,, U^{\otimes L} \right] &= 0 \,, \label{UHam} \\
     \left[ \mathbb{H}(\phi, -\phi)\,, W(0)^{\otimes L} \right] &= 0\,, \label{WHam}
\end{align}
see Eqs. \eqref{U1transf}, \eqref{Z2symmetry}, \eqref{W_Sym2}, respectively. The Hamiltonian also has the additional CP symmetry,
\begin{equation}
    \left[\mathbb{H} \,, V^{\otimes L}\, \Pi  \right] = 0\,,
    \label{CPHam} 
\end{equation}
see Appendix \ref{sec:proofs}. We emphasize that this symmetry does \textbf{not} extend to the full transfer matrix, which has only the CPT symmetry \eqref{CPT}.

The $U(1)$ generators commute with the $Z_2$ symmetry \eqref{Z2symmetry}
\begin{equation}
    \left[ \mathbbm{h}_j \,, U^{\otimes L} \right]  = 0\,, \qquad j = 1\,, 2 \,,
\end{equation}
and they transform into each other under $W(0)$ \eqref{RWsymmetry}
\begin{align}\label{Wgens}
      W(0)^{\otimes L}\, \mathbbm{h}_1 \, W(0)^{\otimes L} &= - \mathbbm{h}_2\,,\nonumber \\
      W(0)^{\otimes L}\, \mathbbm{h}_2 \, W(0)^{\otimes L} &= - \mathbbm{h}_1\,.
\end{align}
Under the CP symmetry, the $U(1)$ generators transform as
\begin{equation}
     V^{\otimes L}\,   \Pi\, \mathbbm{h}_j\, \Pi\, V^{\otimes L} = -\mathbbm{h}_j\,, \qquad j = 1\,, 2\,.
     \label{CPgens}
\end{equation}

The symmetry transformations \eqref{WHam} together with \eqref{CPHam} induce degeneracies in the energy spectrum between different sectors of the $U(1)$-charges.  For the analysis of the finite-size spectrum, it is sufficient to focus on one representative of a given energy level, keeping these degeneracies implicit.  The symmetries \eqref{WHam},  \eqref{CPHam} allow one to restrict to the case   $h_1\ge |h_2|$ for suitably chosen twist angles. In addition, we found by exact diagonalization of the Hamiltonian for small system sizes that one can further restrict to 
\begin{align}
    0\le h_2 \le h_1 \,. 
    \label{h_Constraint}
\end{align}
 Note that all the sectors specified by \eqref{h_Constraint} can be accessed by the above Bethe ansatz, see \eqref{NumBRConstraint} and \eqref{hvals}.  
Further, we should stress that the defined Hamiltonian is non-Hermitian. This leads to complex eigenvalues. On numerical grounds, we find, however, that the energies of the ground state and lowest excitations are real.  In the rest of this work, we study states parameterized by the classes of Bethe root configurations listed in the following section.  These states, too, turn out to have real energies.

\section{Methodology of studying the scaling limit}\label{sec:methodology}

To study the scaling limit of a lattice model, one should define an $L$-dependence to the low-lying energy states. Such an assignment $\ket{\Psi_L}$ will be called an RG-trajectory. For the ground state or for the lowest energy states in the disjoint sectors $(h_1,h_2)$ of the Hilbert space, such an assignment is clear. On the other hand, constructing individual RG trajectories $\ket{\Psi_L}$ for generic low-energy states is not a trivial task. The fact that the considered model is integrable allows for the following strategy: For small initial lattice sizes $L=L_{\rm{in}} \lesssim 6$, we diagonalize the Hamiltonian in the subspace spanned by eigenvectors of the $U(1)$ generators with given eigenvalues  $h_1$, $h_2$.\footnote{Note that the full Hilbert space has dimension $6^{L}$.} The eigenvalues are then matched via (\ref{Energy}) with a solution of the Bethe ansatz equations $\{u^{[1]},u^{[2]}\}^{m_1}_{m_2}$ where $m_1$ and $m_2$ are determined by \eqref{hvals}. The state $\ket{\Psi_{L}}$ at higher $L=L_{\rm{in}}+2$ is obtained by solving the Bethe ansatz equations for a pattern of Bethe roots that qualitatively resembles the one of the state $\ket{\Psi_{L_{\rm{in}}}}$. Via this procedure, we construct the RG trajectory $\ket{\Psi_{L}}$ up to $L \sim 2000$ without relying on a direct diagonalization of the Hamiltonian, which is an impossible task for $L\gg 1$ since the size of the Hilbert space grows exponentially. 

\subsection{Considered class of states}

\begin{figure}[H]
\begin{center}
\scalebox{0.99}{
\begin{tikzpicture}
\node at (0,0) {\includegraphics[width=12cm]{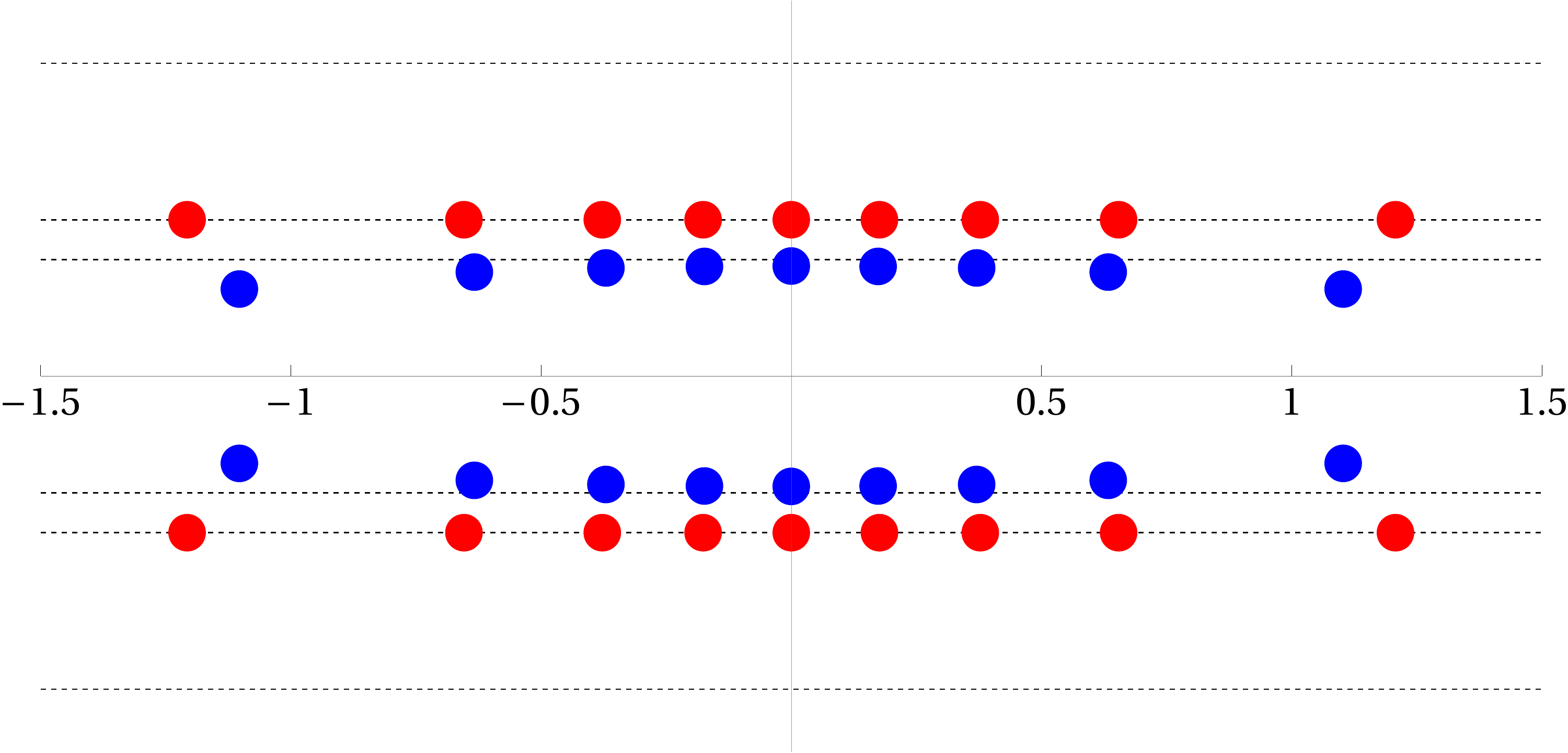}};
\node at (0.00,3.1)   {$\Im m(u)$};
\node at (6.65,0.00)   {$\Re e(u)$};

\node at (6.05,2.4)   {$\pi$};
\node at (6.05,1.2)   {$\frac{\pi}{2}$};
\node at (6.5,0.85)   {$-\gamma+\frac{\pi}{2}$};
\node at (6.1,-1.24)   {$-\frac{\pi}{2}$};
\node at (6.5,-0.85)   {$\gamma-\frac{\pi}{2}$};
\node at (6.1,-2.4)   {$-\pi$};

\end{tikzpicture}
}
\end{center}
\caption{\label{BR_GS}\small
Bethe root configuration of the ground state for $L=18$ and $\gamma=0.4$ plotted in the complex $u$-plane. Blue (red) symbols denote level 1 (2) roots. One can clearly see the pattern (\ref{BR_GS_1}).
}
\end{figure}
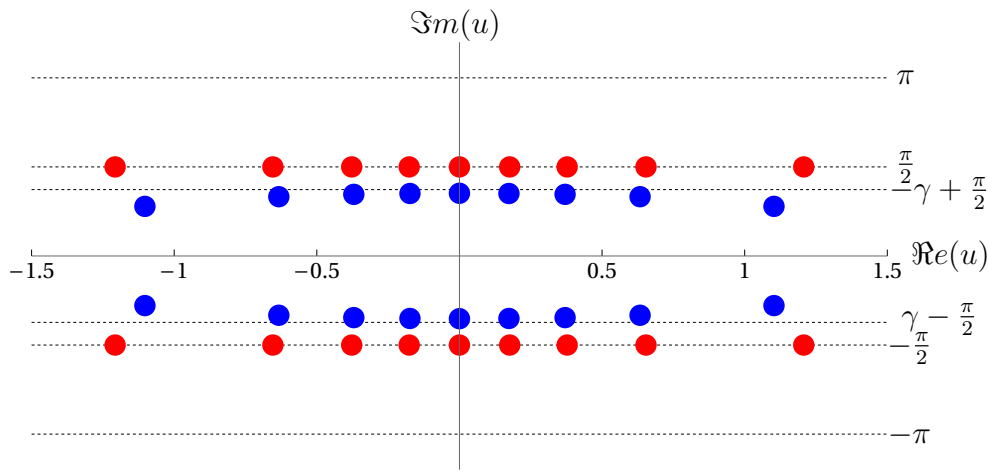

In the above procedure, it is essential to understand the structure of the low-energy spectrum in terms of the Bethe roots. However, the particular structure depends essentially on the domain of the parameters $\gamma$, $\phi_1$ and $\phi_2$. In the regime
\begin{align}
 \gamma \in (0,\tfrac{\pi}{4}) \label{Parameter_Domain}
\end{align}
and for small $\phi_{1,2}$, we found that the bulk of the Bethe root configurations corresponding to low energy states consists of 4-strings, each containing a pair of conjugate roots on both levels centered at real $x_j$: 
\begin{equation}
\label{BR_GS_1}
\begin{aligned}
  &\BR{1}{} \longrightarrow\quad \BRv{1}{j}=x_j+\delta^{[1]}_j+ \frac{\ri\pi}{2}-\ri\gamma-\ri\epsilon^{[1]}_j\,, \quad \BRvbar{1}{j}=x_j+\delta^{[1]}_j- \frac{\ri\pi}{2}+\ri\gamma+\ri\epsilon^{[1]}_j\,, \\
  &\BR{2}{} \longrightarrow  \quad  \BRv{2}{j}=x_j+\delta^{[2]}_j+\frac{\ri \pi}{2}+\ri\epsilon^{[2]}_j\,, \qquad\quad\, \BRvbar{2}{j}=x_j+\delta^{[2]}_j-\frac{\ri \pi}{2}-\ri\epsilon^{[2]}_j \,,
\end{aligned}
\end{equation}
where $j\le \frac{L}{2}$ and $\delta^{[k]}_j,\epsilon^{[k]}_j$ are small real deviations. For even system sizes, the ground state of the system is realized in the sectors $h_1=h_2=0$ with $\epsilon^{[2]}_j\equiv 0 $. See  Fig.~\ref{BR_GS} for the ground state of the $L=18$ chain. The low-energy spectrum is described by various root configurations. In this work, we focus on a particular class of states described by the following additional roots outside these 4-string configurations:
\begin{enumerate}[label=\roman*)]
    \item \label{i} Level-$1$ roots on the line $\frac{\ri \pi}{2}$ 
    \item \label{ii} Level-$2$ roots placed on the line $\ri \pi$. 
    \item \label{iii} Level-$2$ roots placed on the real line. 
\end{enumerate}
subject to the constraint \eqref{NumBRConstraint}.

\section{Root density approach for the ground state}\label{sec:ground}

For even $L$ the ground state is parameterized by roots arranged in the configuration (\ref{BR_GS_1}) where  $j$ runs from one to $\frac{L}{2}$ and $\epsilon^{[2]}_j$ is set to zero. Further, we find numerically that the remaining deviations $\delta^{[k]}_j,\epsilon^{[1]}_j$ in (\ref{BR_GS_1})  tend to zero as $L\to \infty$. Hence, we can study the ground state in the root density approach \cite{YaYa69}. 
By inserting (\ref{BR_GS_1}) into the Bethe equations and taking the logarithm one obtains the following counting function for the real centers: 
\begin{align}
    z^x(x)&=\frac{1}{2\pi} \psi(x,2\gamma)+\frac{1}{2\pi L}\sum^{\frac{L}{2}}_{k=1}\chi(x-x_k,4\gamma)\,,
\end{align}
where
\begin{align}
     \chi(x,y)&=2\arctan\left(\tanh(x)\cot(y)\right)\,,\quad
      \psi(x,y)=2\arctan\left(\tanh(x)\tan(y)\right) \,.
\end{align}
Upon differentiation we obtain the following linear integral equation for the root density defined by $\rho^x(x)=\partial_xz^x(x)$:

\begin{align}
    \rho^x(x)&=\frac{1}{2\pi} \psi'(x,2\gamma)+\frac{1}{2\pi}\int^{\infty}_{-\infty}\mathrm{d }x'\, 
    \chi'(x-x',4\gamma)
    \rho^x(x')\,,
    \label{Lin_Int}
\end{align}
which is solved by Fourier transform giving: 
\begin{align}
\label{rho_thermo}
    \rho^x(x)&=\frac{1}{2(\pi-4\gamma)}\frac{1}{\cosh(\frac{\pi x}{\pi-4\gamma})}\,.
\end{align}
The density is positive and becomes singular at $\gamma=\tfrac{\pi}{4}$, giving additional  support to our choice of the parameter domain (\ref{Parameter_Domain}) for $\gamma$. 
Similarly, the \emph{dressed} energy $\epsilon^x(x)$ of excitations corresponding to  the removal of a four-string is obtained from the same linear integral equation as (\ref{Lin_Int}) but with the driving term $\psi$ replaced by  $\epsilon^x_0(x)=\epsilon_0(x+\frac{\ri \pi}{2}-\ri \gamma)+\epsilon_0(x-\frac{\ri \pi}{2}+\ri \gamma)$, where $\epsilon_0$ has been defined in (\ref{Energy}) above.  For $\gamma$ in the domain (\ref{Parameter_Domain}), these excitations turn out to be gapless with a linear dispersion. The corresponding Fermi velocity is
\begin{align}
    v_F&:= \frac{1}{2\pi }\lim_{\Lambda\to \infty} \frac{1}{\rho^x(\Lambda)}\frac{\mathrm{d}}{\mathrm{d}\Lambda}\epsilon^x(\Lambda)=\frac{\pi \sin(2\gamma)}{\pi-4\gamma }\,. \label{vF}
\end{align}
Finally, using (\ref{rho_thermo}), we obtain the energy density $e_\infty$ in the thermodynamic limit
\begin{align}
    e_\infty&=-\frac{\sin(2\gamma)}{2}\int^{\infty}_{-\infty}\mathrm{d}\omega  \, \frac{ \sinh (2 \gamma  \omega )}{\sinh \left(\frac{\pi  \omega }{2}\right)} \frac{1}{\cosh(\frac{1}{2} (\pi -4 \gamma ) \omega)} \,. \label{einf}
\end{align}

\section{Analysis of the finite-size spectrum}\label{sec:fsize}

As the model is critical, the spectrum of low-energy excitations can be described within the framework of a conformal field theory. In this paper, we extract the central charge and investigate the first features of the underlying CFT, such as the spectrum of scaling dimensions. The following prediction from conformal field theory is expected to hold \cite{BlCN86,Affl86,Card86a},
\begin{align}
   \frac{L}{2\pi v_F}\left(E -Le_\infty\right)&\simeq -\frac{c}{12}+h+\bar{h} \,,\label{Cardy_Formula}
\end{align}
where $e_\infty$ and $v_F$ are given by \eqref{einf} and \eqref{vF}, respectively. Hence, by studying the asymptotic behavior of the energies on the lattice, one can access the central charge $c$ and the conformal weights $h,\,\bar{h}$ of primaries in the  underlying CFT. Note that on the right-hand side of (\ref{Cardy_Formula}), only the sum of the central charge and the scaling dimension $X=h+\bar{h}$ appears, such that one cannot determine $c$ or $h,\, \bar{h}$ on their own. Suitable measures of the scaling dimensions and the central charge are given by the \emph{effective} scaling dimensions $X_{\mathrm{eff}}$ and the \emph{effective} central charge $c_{\mathrm{eff}}$ defined by
\begin{align}
   X_{\mathrm{eff}}&= \frac{L}{2\pi v_F}\left(E -Le_\infty\right)\label{X_eff_def} \,, \\
   c_{\mathrm{eff}}&=-\frac{6L}{\pi v_F}\left(E_{GS} -Le_\infty\right)\label{c_eff_def} \,.
\end{align}
The quantity $E_{GS}$ in (\ref{c_eff_def}) is the ground state energy.

Besides (\ref{Cardy_Formula}) we have also the following relation between the scaling dimensions and the eigenvalue of one-site translation operator \eqref{onesiteshift}
\begin{align}
       \re^{\ri \, P}&=\re^{\frac{2\ri \pi}{L}\left(h-\bar{h}\right)}\,.\label{Conformal Spin}
\end{align}
One can easily show that 
\begin{align}
    \re^{\ri \, \mathbb{P}\, L}=\exp{\left\{ 2\ri \pi \left( \mathbbm{h}_1 \, \frac{\phi_1}{2\pi}+\mathbbm{h}_2\,\frac{\phi_2}{2\pi}\right)\right\}}\,.
\end{align}
Comparing with (\ref{Conformal Spin}), one analytically obtains the result for the difference $h-\bar{h}$
\begin{align}
    h-\bar{h}=h_1 \,\frac{\phi_1}{2\pi}+h_2\,\frac{\phi_2}{2\pi} \mod 1\,.\label{h-barh}
\end{align}
To proceed further in our analysis, we relied on numerical methods whose results we present in the following sections.

\subsection{Compact part}

\begin{figure}[H]
    \centering
    \begin{minipage}[b]{.49\linewidth}
    \scalebox{0.93}{
\begin{tikzpicture}
\node at (0,0) {\includegraphics[width=0.9\linewidth]{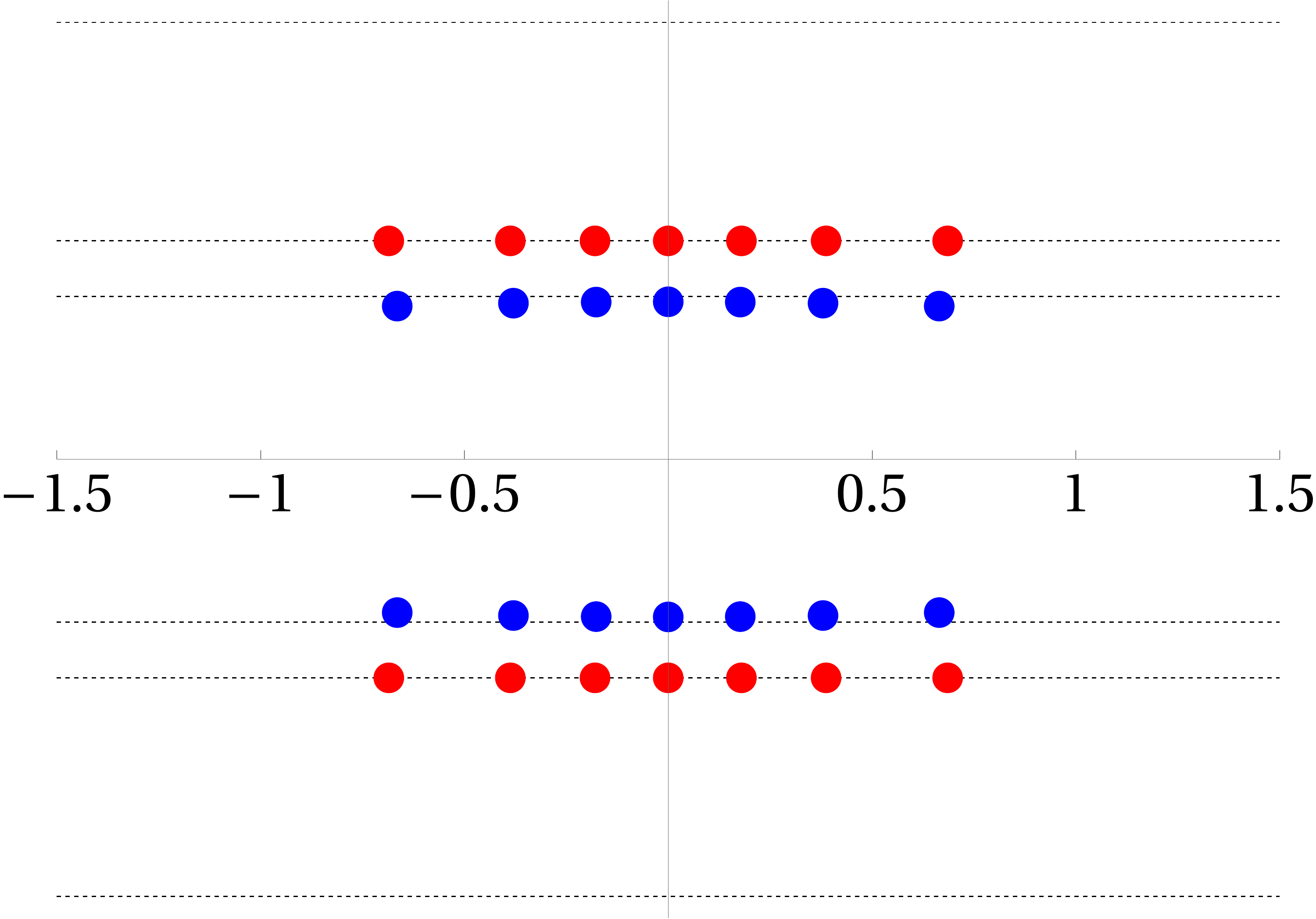}};
\node at (0.00,2.9)   {\small$\Im m(u)$};
\node[anchor=west] at  (3.75,0.00)  {\small$\Re e(u)$};
\node[anchor=west] at (3.5,2.5)   {\small$\pi$};
\node[anchor=west] at (3.5,1.35)   {\small$\frac{\pi}{2}$};
\node[anchor=west] at (3.5,0.85)   {\small$\frac{\pi}{2}-\gamma$};
\node[anchor=west] at (3.5,-0.85)   {\small$-\frac{\pi}{2}$};
\node[anchor=west] at (3.5,-1.35)  {\small$-\frac{\pi}{2}+\gamma$};
\node[anchor=west] at (3.5,-2.5)   {\small$-\pi$};

\end{tikzpicture}
}
 \end{minipage}
 \begin{minipage}[b]{.49\linewidth}
      \scalebox{0.93}{
\begin{tikzpicture}
\node at (0,0) {\includegraphics[width=0.9\linewidth]{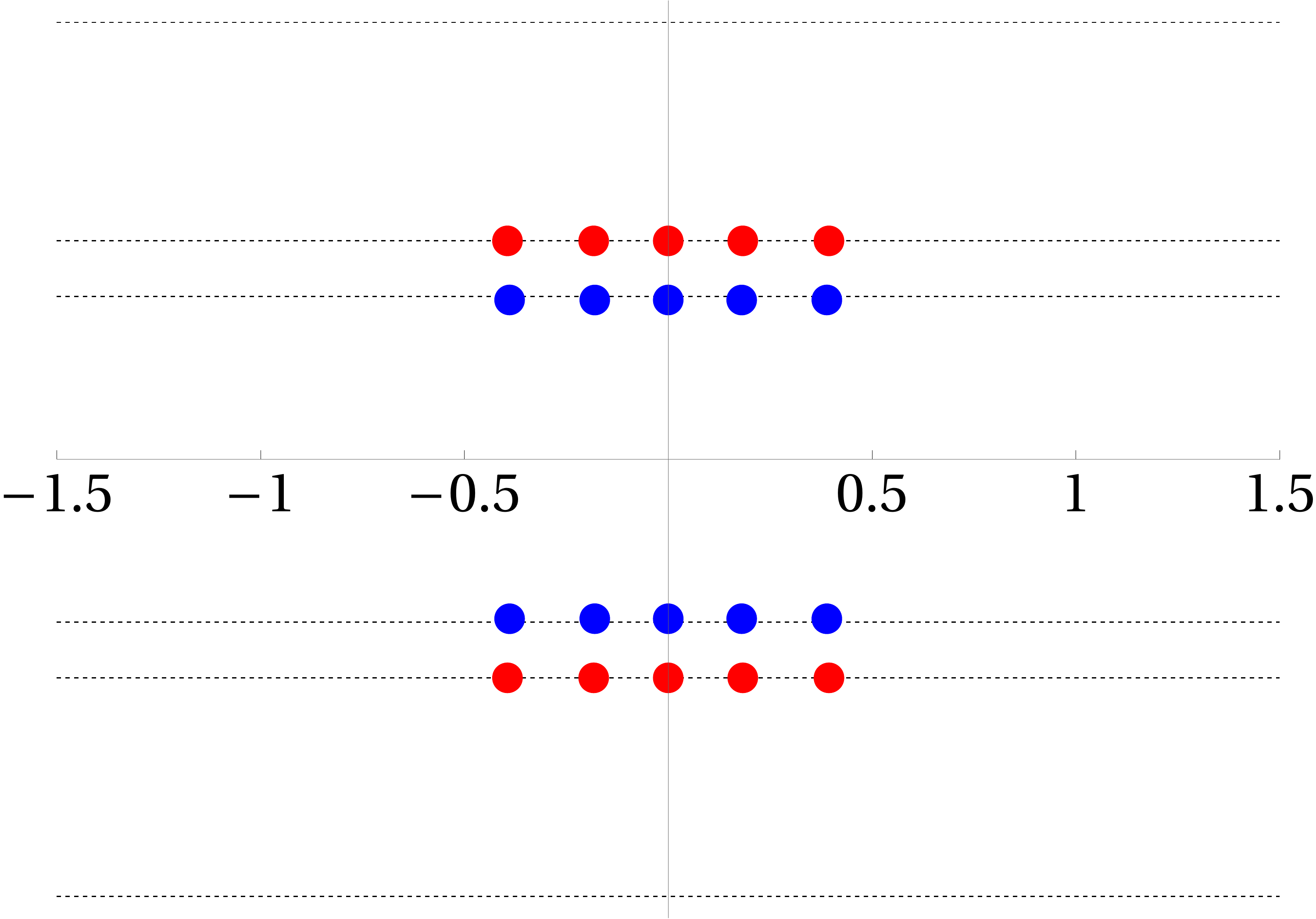}};
\node at (0.00,2.9)   {\small$\Im m(u)$};
\node[anchor=west] at  (3.75,0.00)  {\small$\Re e(u)$};
\node[anchor=west] at (3.5,2.5)   {\small$\pi$};
\node[anchor=west] at (3.5,1.35)   {\small$\frac{\pi}{2}$};
\node[anchor=west] at (3.5,0.85)   {\small$\frac{\pi}{2}-\gamma$};
\node[anchor=west] at (3.5,-0.85)   {\small$-\frac{\pi}{2}$};
\node[anchor=west] at (3.5,-1.35)  {\small$-\frac{\pi}{2}+\gamma$};
\node[anchor=west] at (3.5,-2.5)   {\small$-\pi$};
\end{tikzpicture}
}
\end{minipage}
    \caption{Left (right) plot displays the Bethe-root configuration in the complex $u$-plane of an excited state for $L=18$, $\gamma=0.4, \phi_{1,2}=0$ in the sector $h_1=4$ ($8$). Blue (red) symbols denote level 1 (2) roots. This excitation corresponds to removing $2$ ($4$) four-strings from the configuration of the ground state, see Fig.~\ref{BR_GS}. }
    \label{BR_Mode_1}
\end{figure}
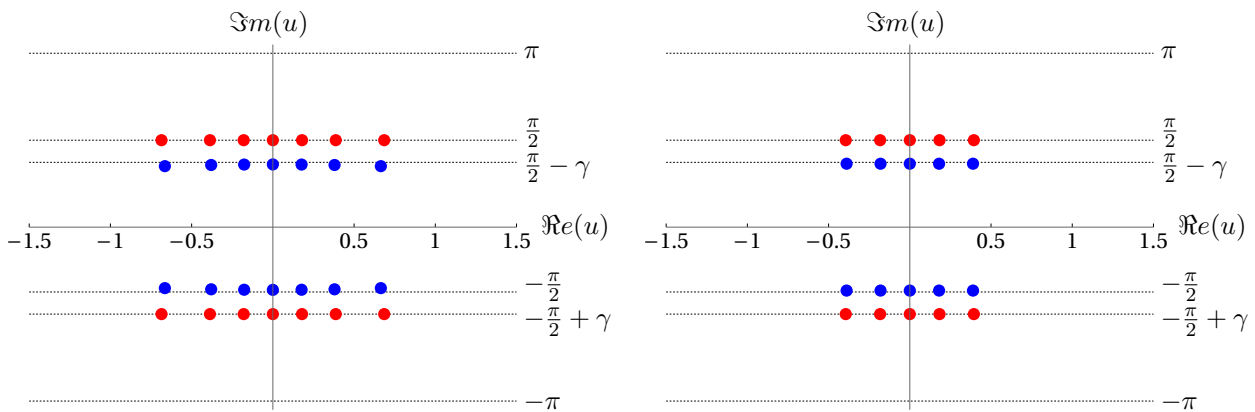

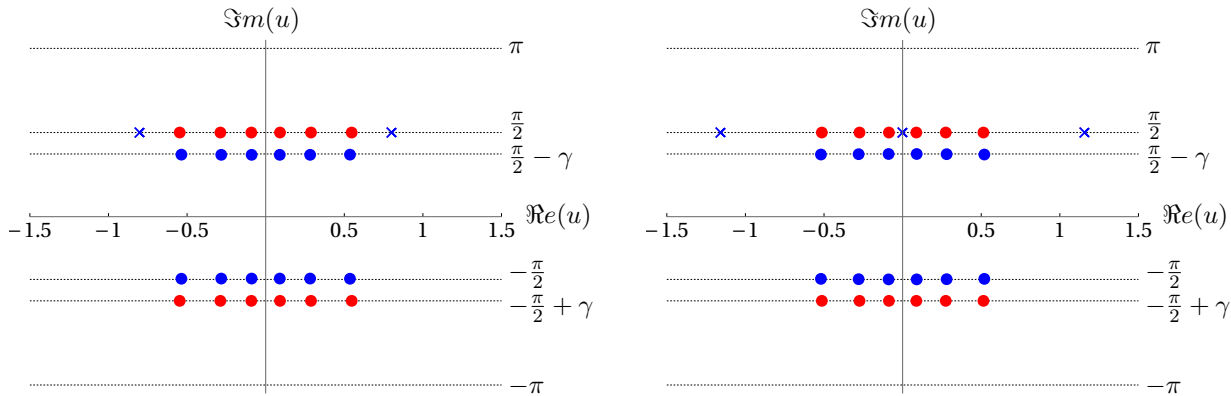
\begin{figure}[H]
    \centering
    \begin{minipage}[b]{.49\linewidth}
    \scalebox{0.9}{
\begin{tikzpicture}
\node at (0,0) {\includegraphics[width=0.9\linewidth]{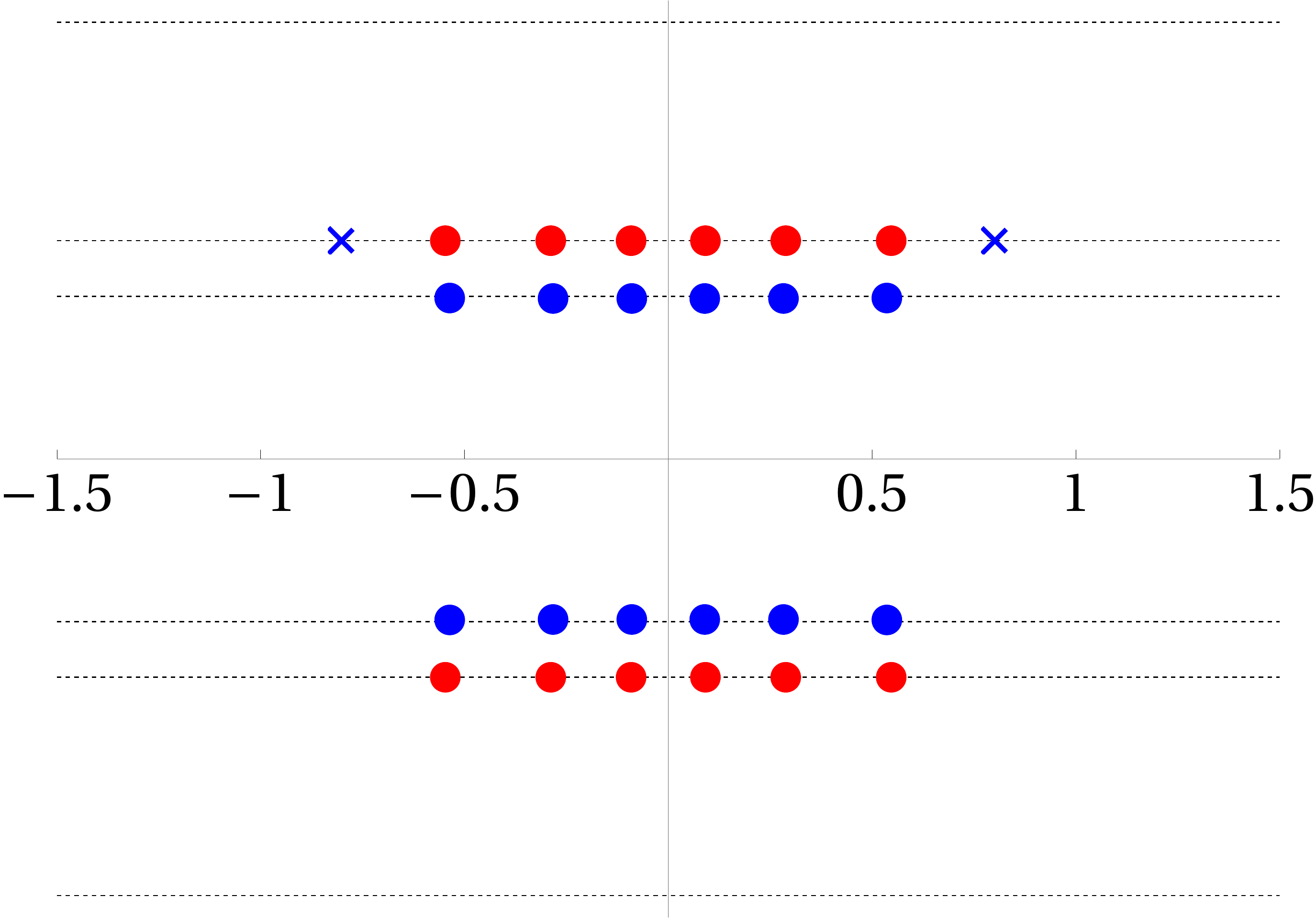}};
\node at (0.00,2.9)   {\small$\Im m(u)$};
\node[anchor=west] at  (3.75,0.00)  {\small$\Re e(u)$};
\node[anchor=west] at (3.5,2.5)   {\small$\pi$};
\node[anchor=west] at (3.5,1.35)   {\small$\frac{\pi}{2}$};
\node[anchor=west] at (3.5,0.85)   {\small$\frac{\pi}{2}-\gamma$};
\node[anchor=west] at (3.5,-0.85)   {\small$-\frac{\pi}{2}$};
\node[anchor=west] at (3.5,-1.35)  {\small$-\frac{\pi}{2}+\gamma$};
\node[anchor=west] at (3.5,-2.5)   {\small$-\pi$};

\end{tikzpicture}
}
 \end{minipage}
 \begin{minipage}[b]{.49\linewidth}
      \scalebox{0.9}{
\begin{tikzpicture}
\node at (0,0) {\includegraphics[width=0.9\linewidth]{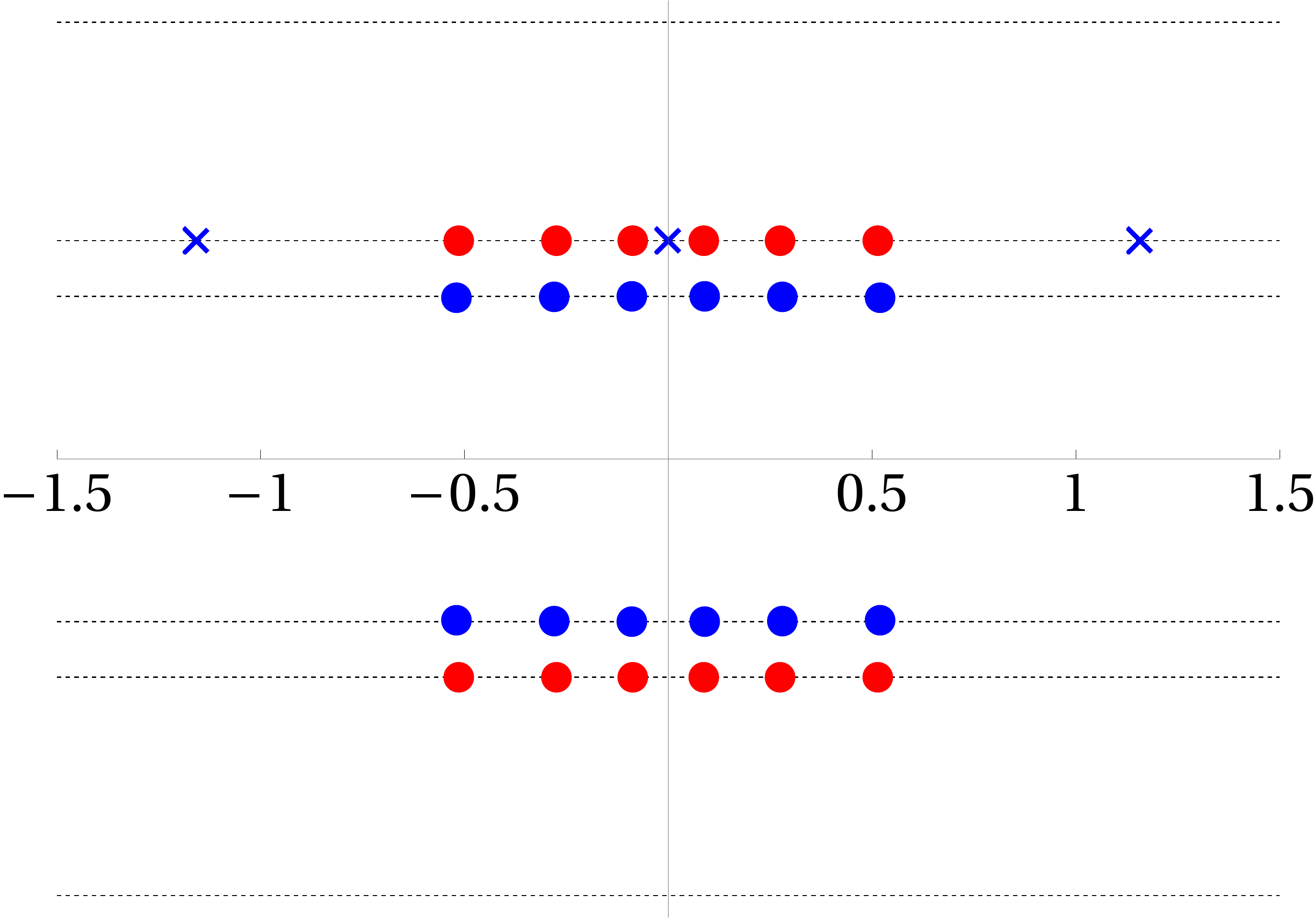}};
\node at (0.00,2.9)   {\small$\Im m(u)$};
\node[anchor=west] at  (3.75,0.00)  {\small$\Re e(u)$};
\node[anchor=west] at (3.5,2.5)   {\small$\pi$};
\node[anchor=west] at (3.5,1.35)   {\small$\frac{\pi}{2}$};
\node[anchor=west] at (3.5,0.85)   {\small$\frac{\pi}{2}-\gamma$};
\node[anchor=west] at (3.5,-0.85)   {\small$-\frac{\pi}{2}$};
\node[anchor=west] at (3.5,-1.35)  {\small$-\frac{\pi}{2}+\gamma$};
\node[anchor=west] at (3.5,-2.5)   {\small$-\pi$};
\end{tikzpicture}
}
\end{minipage}
    \caption{Left (right) plot displays the Bethe-root configuration in the complex plane of an excited state for $L=18,(19)$, $\gamma=0.4, \phi_{1,2}=0$ in the sector $ h_1=4$ and $ h_2=2,(3)$. Blue (red) symbols denote level 1 (2) roots. This excitation is built by placing  2(3) level-$1$ roots ($\times$) on the line $\frac{\ri \pi }{2}$ in addition to the bulk roots ($\bullet$). }
    \label{BR_Second_Compact}
\end{figure}

\begin{figure}[H]
    \centering
    \begin{minipage}[b]{.49\linewidth}
     \includegraphics[width=\linewidth]{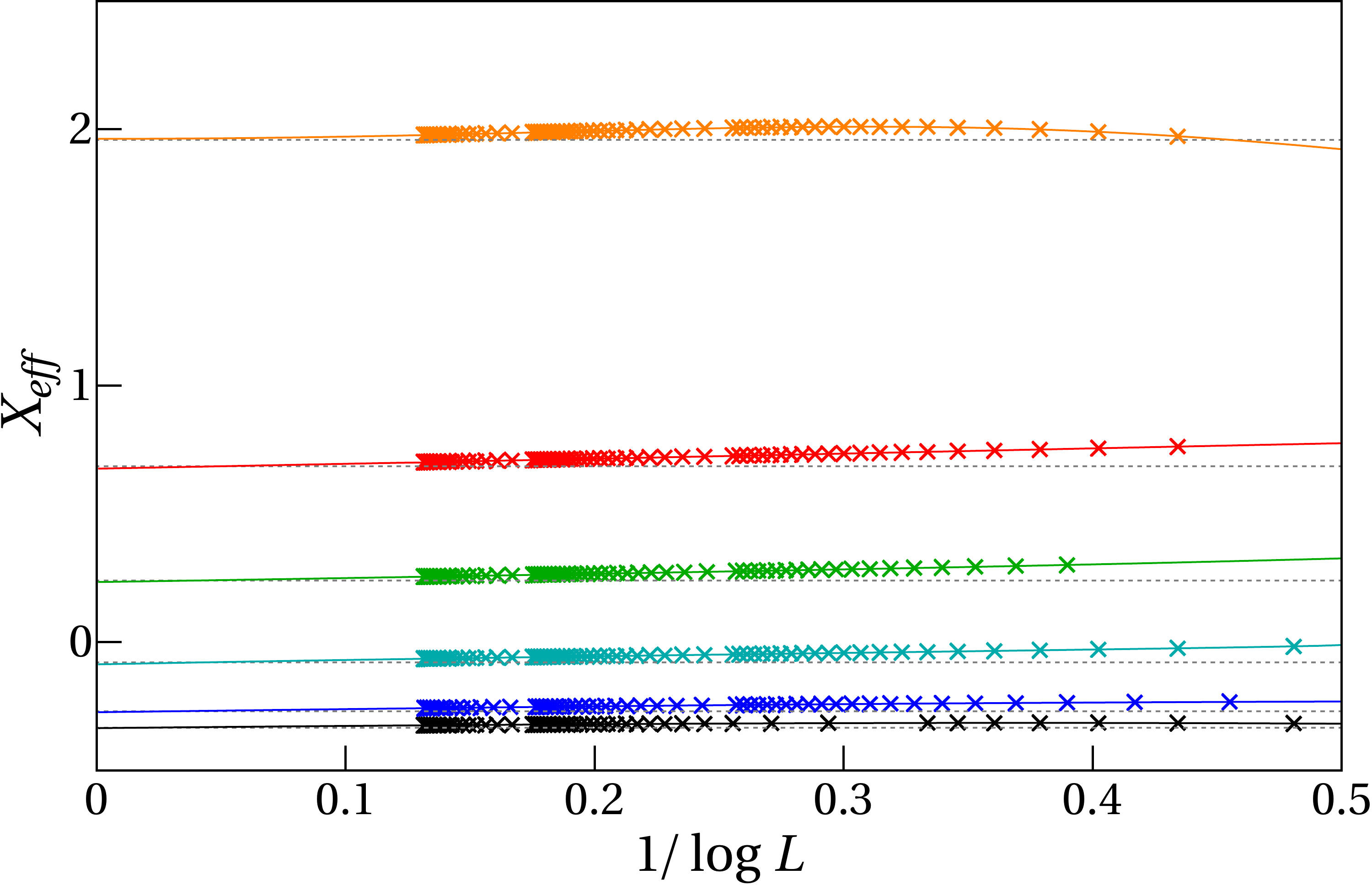}
    \end{minipage} \begin{minipage}[b]{.49\linewidth}
     \includegraphics[width=\linewidth]{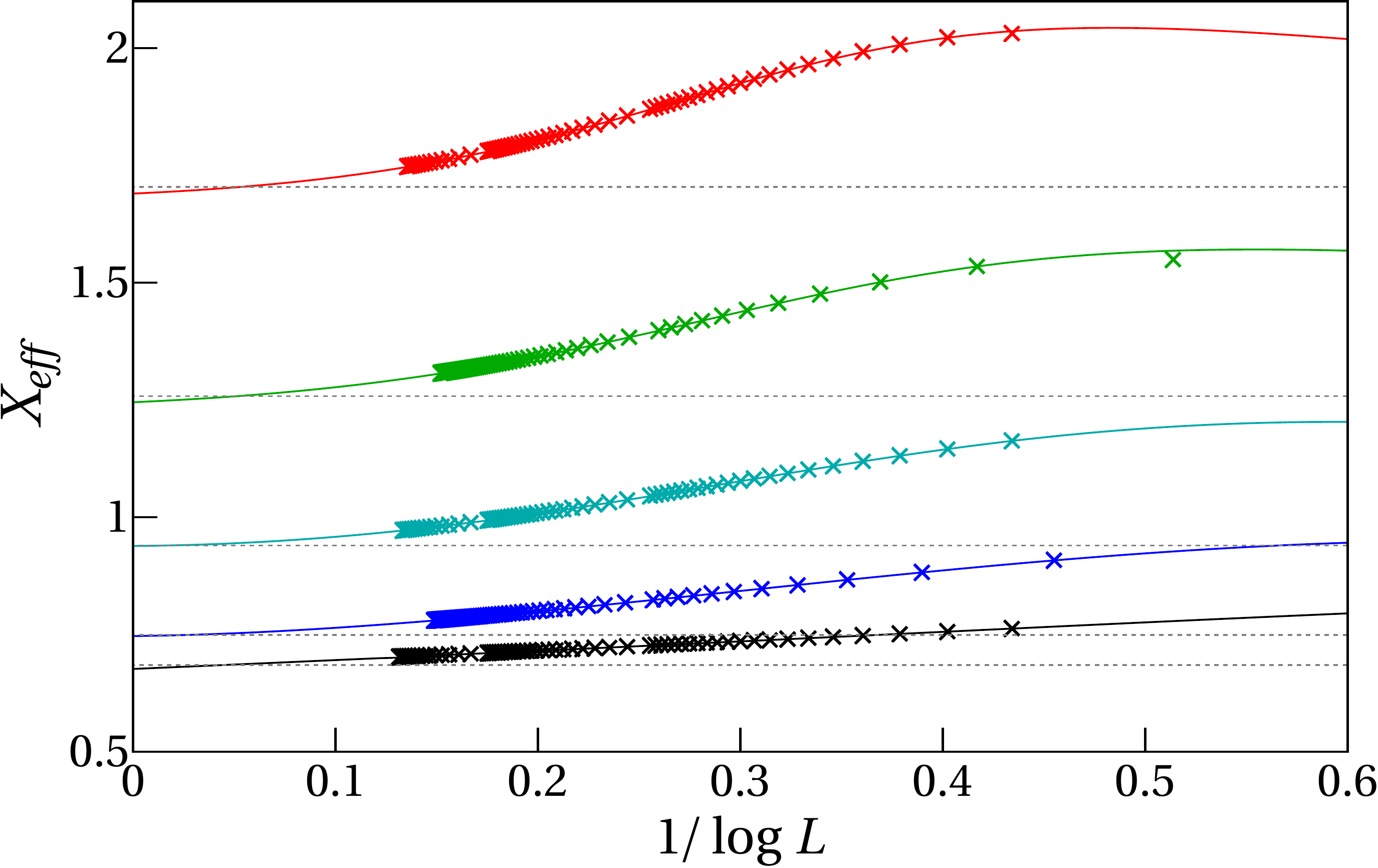}
    \end{minipage}
    \par
       \begin{minipage}[t]{.49\linewidth}
        \caption{Finite-size scaling up to $L\sim 2000$ of the ground state (black) and states with pattern (see Fig.~\ref{BR_Mode_1}) similar to the ground state with $h_2=0$ but with different $U(1)$-charge $\mathbbm{h}_1$, i.e., $h_1=1,2,3,4,6$ in increasing order from below. The crosses are the numerically obtained effective scaling dimensions. The dashed lines are given by formula (\ref{Xeff_Compact_PBC}) with the associated $h_{1,2}$. Solid lines are given by a rational extrapolation. Here $\gamma=0.4$ and $\phi_{1,2}=0$.}\label{Xeff_1_Compact}
    \end{minipage}\rulesep
    \begin{minipage}[t]{.49\linewidth}
        \caption{Finite-size scaling up to $L\sim 2000$  for states with similar root configuration, as depicted in Fig.~\ref{BR_Second_Compact}. Crosses are the numerically obtained effective scaling dimensions. The $U(1)$-charges $(h_1,h_2)$ take the values $(4,0),(4,1),(4,2),(4,3),(4,4)$ labelled from below. Dashed lines are given by equation (\ref{Xeff_Compact_PBC}) and the solid lines are given by a rational extrapolation. Here $\gamma=0.4$ and $\phi_{1,2}=0$.\label{Xeff_2_Compact}}
    \end{minipage}
  
    \label{Compact_FS_Plot}
\end{figure}

In this section, we investigate two classes of fundamental excitation patterns. In terms of the Bethe roots, the first class is simply built from configurations following the structure (\ref{BR_GS_1}) but with a non-zero $h_1$ in contrast to the ground state. Here, the eigenvalue $h_2$ of the $U(1)$-charge $\mathbbm{h}_2$ is kept the same as for the ground state, i.e. $h_2=0$. See Fig.~\ref{BR_Mode_1} for an illustration.\\\\
The second class of excitations gives $h_2$ a non-vanishing value. This is accomplished on the level of the Bethe roots by mechanism (\ref{i}, i.e.\ by placing additional level-1 roots on the line $\frac{\ri \pi}{2}$.  We have illustrated this type of excitations in Fig.~\ref{BR_Second_Compact}.
We start our numerical analysis by investigating the scaling behavior of the ground state. We obtain
\begin{align}
    c_{\rm{eff}}=4\,.\label{C_eff_N}
\end{align}
Further, for periodic boundary condition $\phi_{1,2}=0$, we have constructed the RG trajectories for various excited states based on the mechanisms discussed above.  Exemplary plots of the numerical data for finite $L$ calculated by the Bethe ansatz and their extrapolations to $L\to \infty$ are given in Figs.~\ref{Xeff_1_Compact}-\ref{Xeff_2_Compact}. Here, the extrapolation procedure is based on the assumption that the effective scaling dimensions are rational functions of $\frac{1}{\log(L)}$.  We conclude that they flow to the following effective scaling dimensions:
\begin{align}
    X^{\rm{Com}}_{\mathrm{eff}}=-\frac{4}{12}+\frac{(h_1     )^2}{2 \,\kk}
     +\frac{( h_2 )^2}{2 \,\kk} \,.\label{Xeff_Compact_PBC}
\end{align}
In (\ref{Xeff_Compact_PBC}) the first term accounts for the effective central charge and agrees with our findings (\ref{C_eff_N}).  Note the exchange symmetry of $h_1 $ and $h_2$ observed in the spectrum. The parameter $\kk$ specifying the amplitudes is related to the anisotropy by 
\begin{align}
    \kk=\frac{\pi}{\gamma}\,.
\end{align}

\subsubsection{Spectrum flow of the compact modes under twists}

We now turn to the extension of the formula \eqref{Xeff_Compact_PBC} to small non-vanishing twist angles. The analytic expression (\ref{h-barh}) and symmetry arguments suggest the following generalization
\begin{align}
    X^{\rm{Com}}_{\rm{eff}}(h_1,h_2,\phi_1,\phi_2)=& -\frac{4}{12}+\frac{(h_1+           \kk \,   \frac{\phi_1}{2\pi}      )^2}{4 \,\kk}
     +\frac{( h_2+           \kk \,   \frac{\phi_2}{2\pi}        )^2}{4 \,\kk}+\frac{( h_1-           \kk \,   \frac{\phi_1}{2\pi}      )^2}{4\, \kk}+\frac{( h_2-           \kk \,  \frac{\phi_2}{2\pi}      )^2}{4\, \kk}\,.\label{Xeff_Compact_Twist}
\end{align}
We have numerically verified the above expression by using the data of the periodic model by applying the following iterative method: We start with a solution $\{u^{[1]},u^{[2]}\}^{\phi^ { \rm{in}}_1}_{\phi^ {\rm{in}}_2}$ of the BAE \eqref{Twisted_BAE} in logarithmic form with a particular
initial set $(\phi^{ \rm{in}}_1,\phi^{ \rm{in}}_{2})$ of twist
values  e.g., $(\phi^{ \rm{in}}_1,\phi^{ \rm{in}}_{2})=(0,0)$. We see that the maximal error using $\{u^{[1]},u^{[2]}\}^{\phi^{ \rm{in}}_1}_{\phi^{ \rm{in}}_2}$ as an initial approximation for the BAE for new values $(\phi^{ \rm{in}}_1+\Delta \phi_1,\phi^{ \rm{in}}_{2}+\Delta \phi_2)$ behave as $\max\left\{|\Delta\phi_1|,|\Delta\phi_2-\Delta\phi_1|\right\}$. Hence, by taking the steps sizes  $\Delta \phi_1,\Delta \phi_2$ small enough, we can iteratively obtain the state at some $(\phi^{ \rm{end}}_1,\phi^{ \rm{end}}_{2})$.

Note that the above form of the effective scaling dimensions is compatible with the symmetries \eqref{W_Sym2},  \eqref{CPHam}. To interpret these results further, consider the conformal weights of a twisted free boson given by \cite{DiFr97}
\begin{align}
    h_{n,\omega}=\frac{1}{2}\left(\frac{n}{2R}+R (\omega+\varphi)\right)^2 \qquad  \bar{h}_{n,\omega}=\frac{1}{2}\left(\frac{n}{2R}- R (\omega+\varphi)\right)^2
    \label{com_boson_cft}
\end{align}
where the integers $n, \omega$ label charge and winding while $\varphi$ parameterizes the twisted boundary condition. By comparing \eqref{com_boson_cft} and \eqref{Xeff_Compact_Twist} we see that the excitations \eqref{Xeff_Compact_Twist} mimic two independent twisted compact bosonic modes with the same compactification radii $R_{1,2}=\sqrt{\frac{\kk}{2}}$ with  charges $n_{1,2}=h_{1,2}$ and zero windings. Despite extensive study of root patterns of the low-lying excitations, we have not been able to identify any state with non-zero winding. Note that the functional dependence on the compactification radii induced by non-zero twists $\phi_{1,2}$ is exactly the same as for two compact bosons as expected from the symmetries of the model.

We want to end this section by the following important remark. The above expressions for the scaling dimensions capture the leading finite-size behavior only. Corrections to (\ref{Cardy_Formula}) can arise, e.g.,\ due to perturbations of the fixed-point Hamiltonian by terms involving irrelevant operators present in the lattice model (\ref{eq:Hamil}) \cite{Card86a}. In the presence of a marginally irrelevant operator, one expects these subleading corrections to contain logarithms \cite{Card86c}.  In the present case we observe such corrections, see e.g.,\ Figs.~\ref{Xeff_1_Compact} and \ref{Xeff_2_Compact}.  As we will argue below, however, these are also a signature of non-compact degrees of freedom in the effective theory describing the critical behavior.

\subsection{Continuous part of the spectrum}

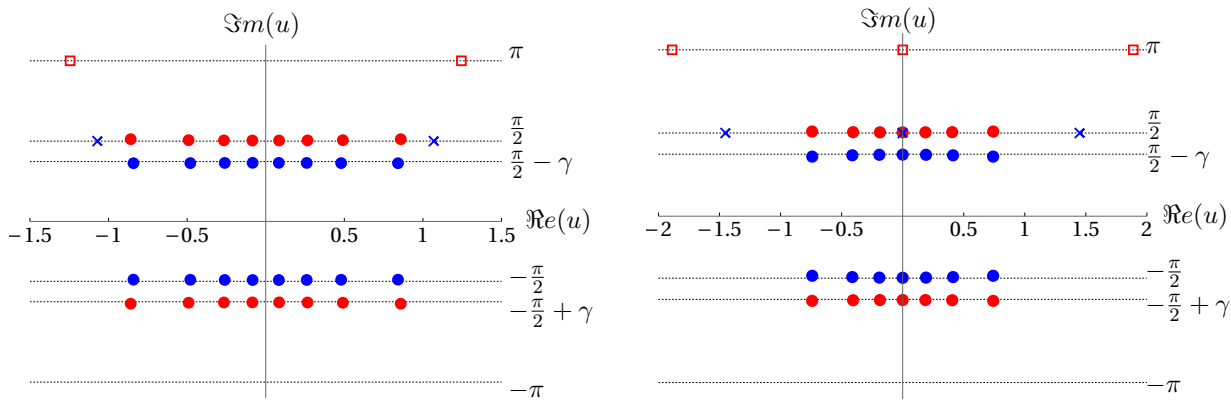
\begin{figure}[H]
    \centering
    \begin{minipage}[b]{.49\linewidth}
    \scalebox{0.9}{
\begin{tikzpicture}
\node at (0,0) {\includegraphics[width=0.9\linewidth]{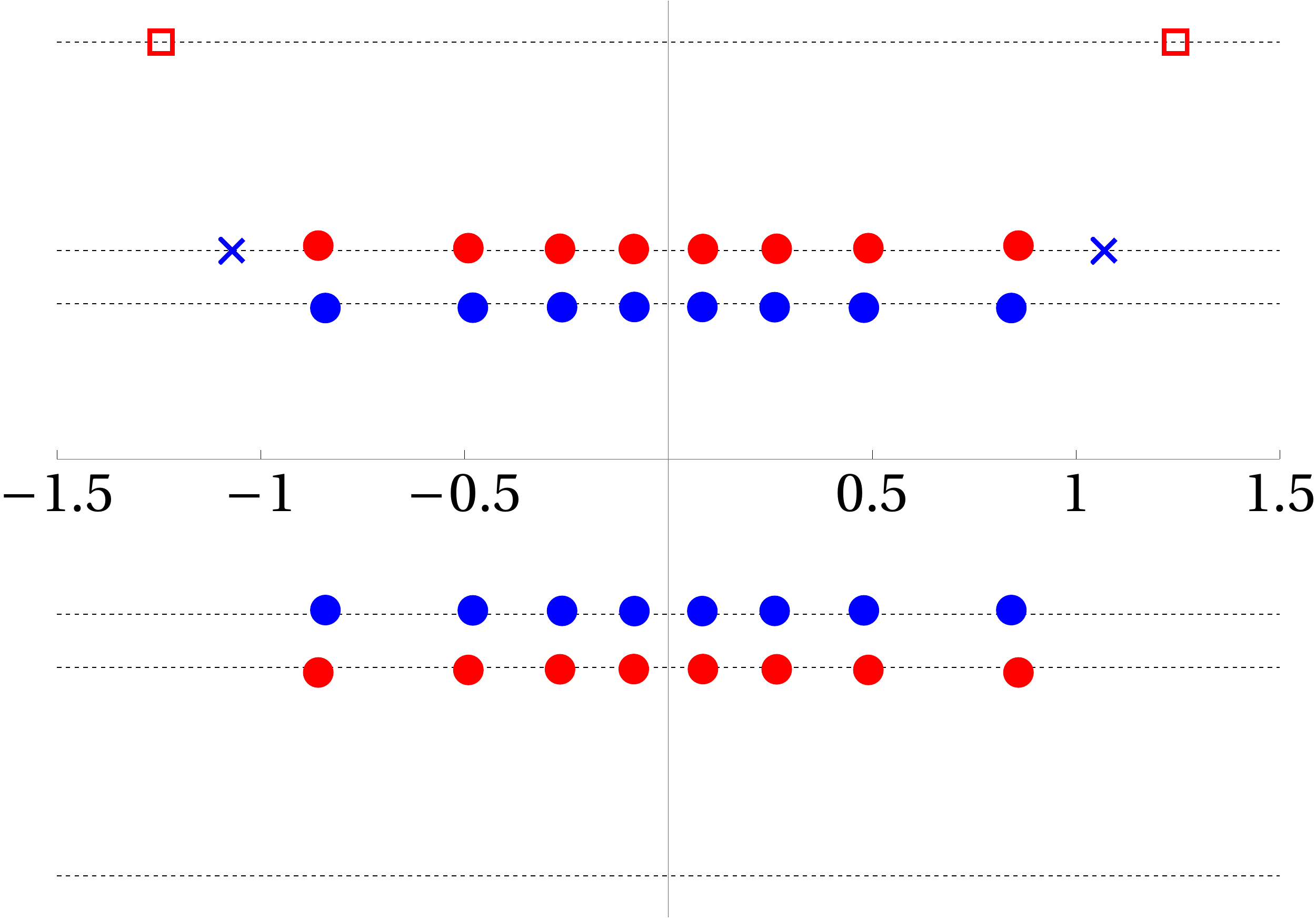}};
\node at (0.00,2.9)   {\small$\Im m(u)$};
\node[anchor=west] at  (3.75,0.00)  {\small$\Re e(u)$};
\node[anchor=west] at (3.5,2.5)   {\small$\pi$};
\node[anchor=west] at (3.5,1.35)   {\small$\frac{\pi}{2}$};
\node[anchor=west] at (3.5,0.85)   {\small$\frac{\pi}{2}-\gamma$};
\node[anchor=west] at (3.5,-0.85)   {\small$-\frac{\pi}{2}$};
\node[anchor=west] at (3.5,-1.35)  {\small$-\frac{\pi}{2}+\gamma$};
\node[anchor=west] at (3.5,-2.5)   {\small$-\pi$};

\end{tikzpicture}
}
 \end{minipage}
 \begin{minipage}[b]{.49\linewidth}
      \scalebox{0.9}{
\begin{tikzpicture}
\node at (0,0) {\includegraphics[width=0.9\linewidth]{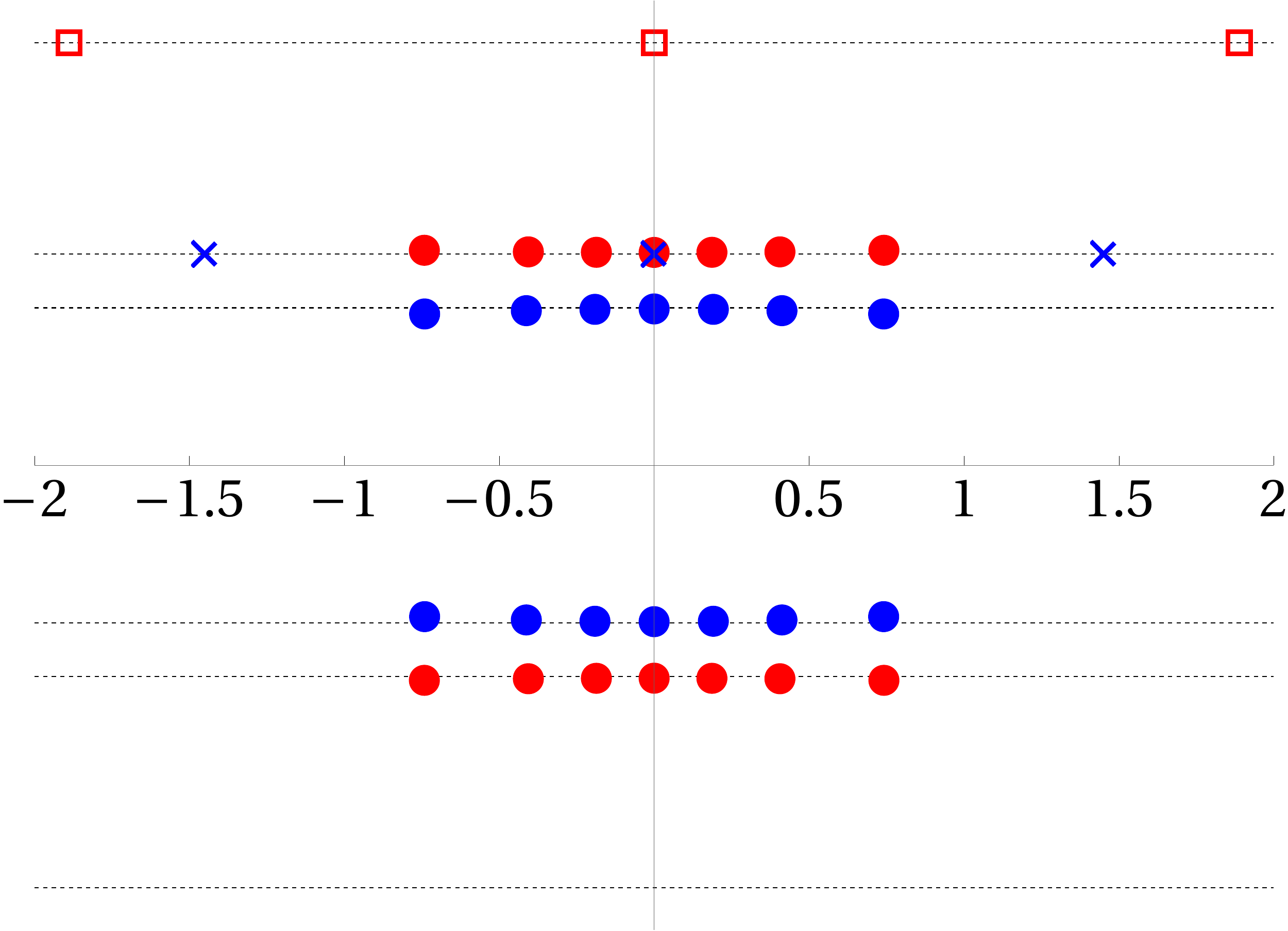}};
\node at (0.00,2.9)   {\small$\Im m(u)$};
\node[anchor=west] at  (3.75,0.00)  {\small$\Re e(u)$};
\node[anchor=west] at (3.5,2.5)   {\small$\pi$};
\node[anchor=west] at (3.5,1.35)   {\small$\frac{\pi}{2}$};
\node[anchor=west] at (3.5,0.85)   {\small$\frac{\pi}{2}-\gamma$};
\node[anchor=west] at (3.5,-0.85)   {\small$-\frac{\pi}{2}$};
\node[anchor=west] at (3.5,-1.35)  {\small$-\frac{\pi}{2}+\gamma$};
\node[anchor=west] at (3.5,-2.5)   {\small$-\pi$};
\end{tikzpicture}
}
\end{minipage}
    \caption{Left (right) plot displays the Bethe-root configuration in the complex plane of an excited state for $L=19,(18)$, $\gamma=0.4$ in the sector $ h_1=1$ and $ h_2=0$. Blue (red) symbols denote level 1 (2) roots. This excitation is built by placing  2(3) level $1$ roots ($\times$) on the line $\frac{\ri \pi }{2}$ and 2,(3) level-2 roots ($\square$) on the line $\ri \pi$ in addition to the bulk roots ($\bullet$). Further, one (and a half four-string) has been removed with respect to  the lowest energy state configuration in this sector.}
    \label{BR_First_Conti}
\end{figure}

Interestingly, there also exist excitations whose scaling dimensions coincide with compact ones \eqref{Xeff_Compact_Twist} up to logarithmic corrections. These excitations can be characterized by the presence of roots of type (\ref{ii} and (\ref{iii}. Consider first the ones of type (\ref{ii}. Examples of their root configurations are displayed in Fig.~\ref{BR_First_Conti}. 

\begin{figure}[H]
    \centering
    \begin{minipage}[b]{.49\linewidth}
     \includegraphics[width=\linewidth]{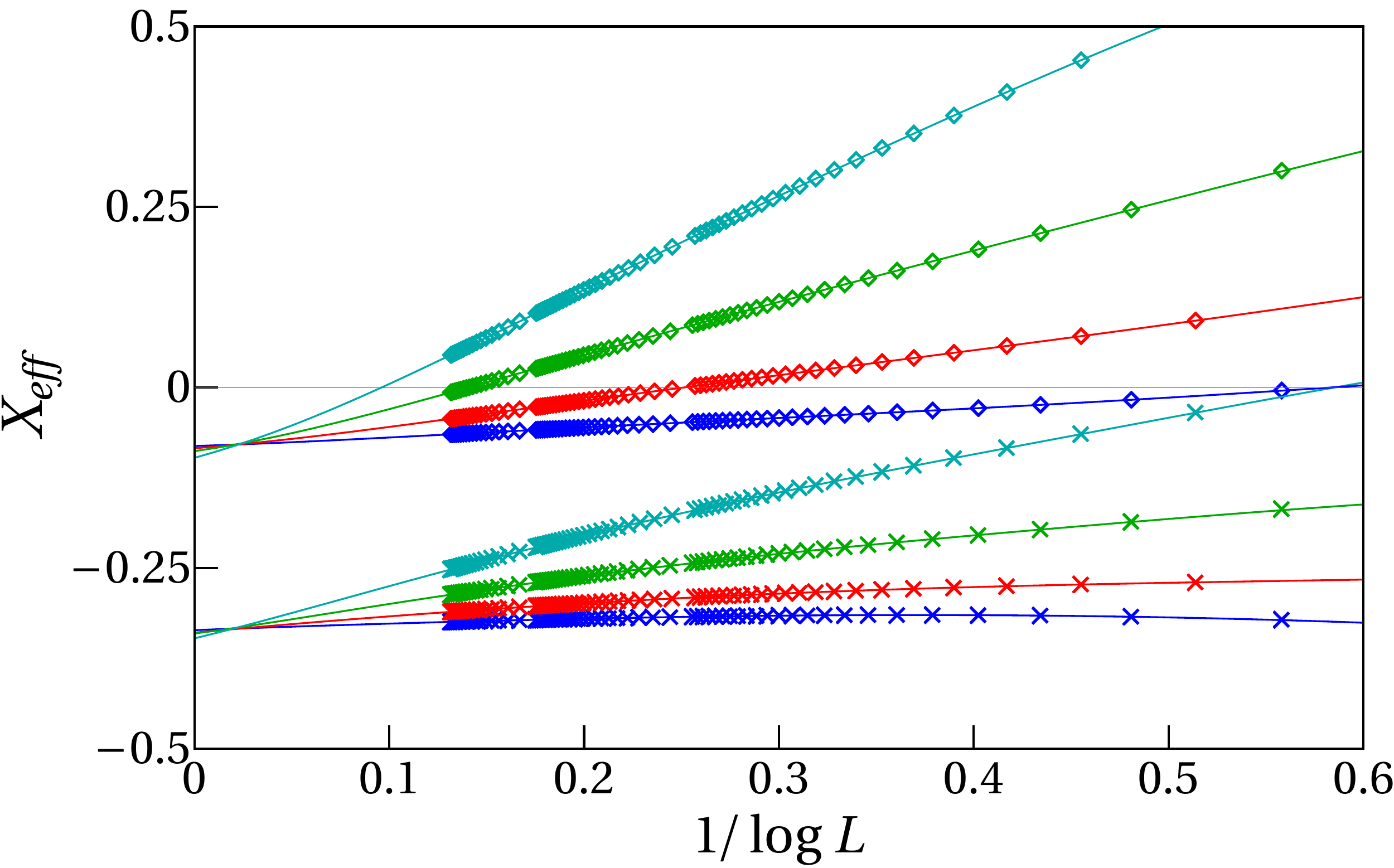}
    \end{minipage} \begin{minipage}[b]{.49\linewidth}
     \includegraphics[width=\linewidth]{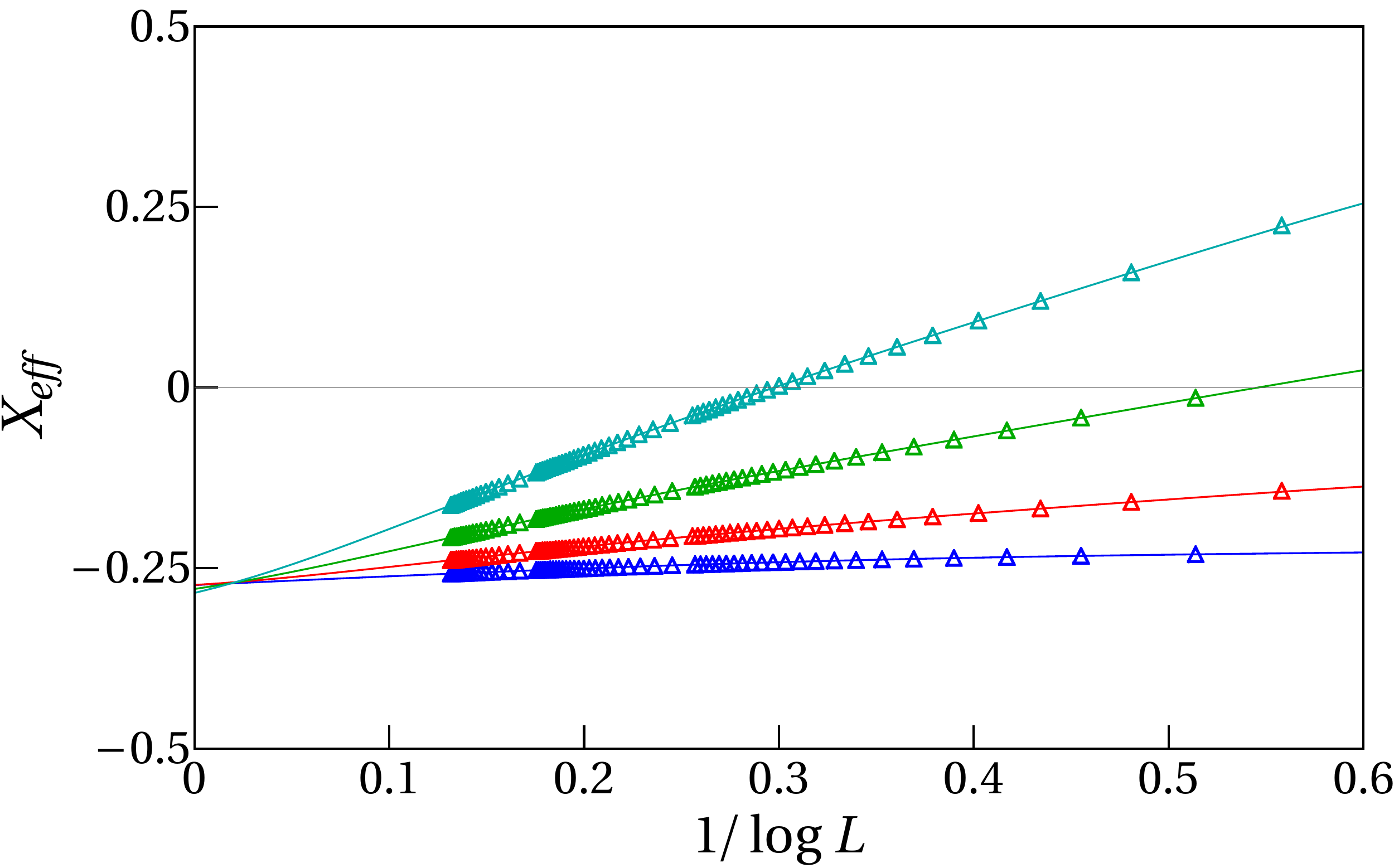}
    \end{minipage}
    \caption{ Left (right) plot displays the finite-Size Scaling up to $L \sim 2000$ in the sectors $(h_1,h_2)$: $(0,0)$ $\times$ , $(2,0)$  $\diamond$ ($(1,0)$ $\triangle$) for states with $M_\pi=0,1,2,3$ in increasing order from below (blue, red, green, cyan). Solid lines are rational extrapolation. One can see clearly the logarithmic dependence of the scaling dimensions. The parameters are set to $\gamma=0.4$ and $\phi_{1,2}=0$.}
    \label{Conti_1_FS_Plot}
\end{figure}

\begin{figure}[H]
    \centering
    \begin{minipage}[b]{.49\linewidth}
    \scalebox{0.9}{
\begin{tikzpicture}
\node at (0,0) {\includegraphics[width=0.9\linewidth]{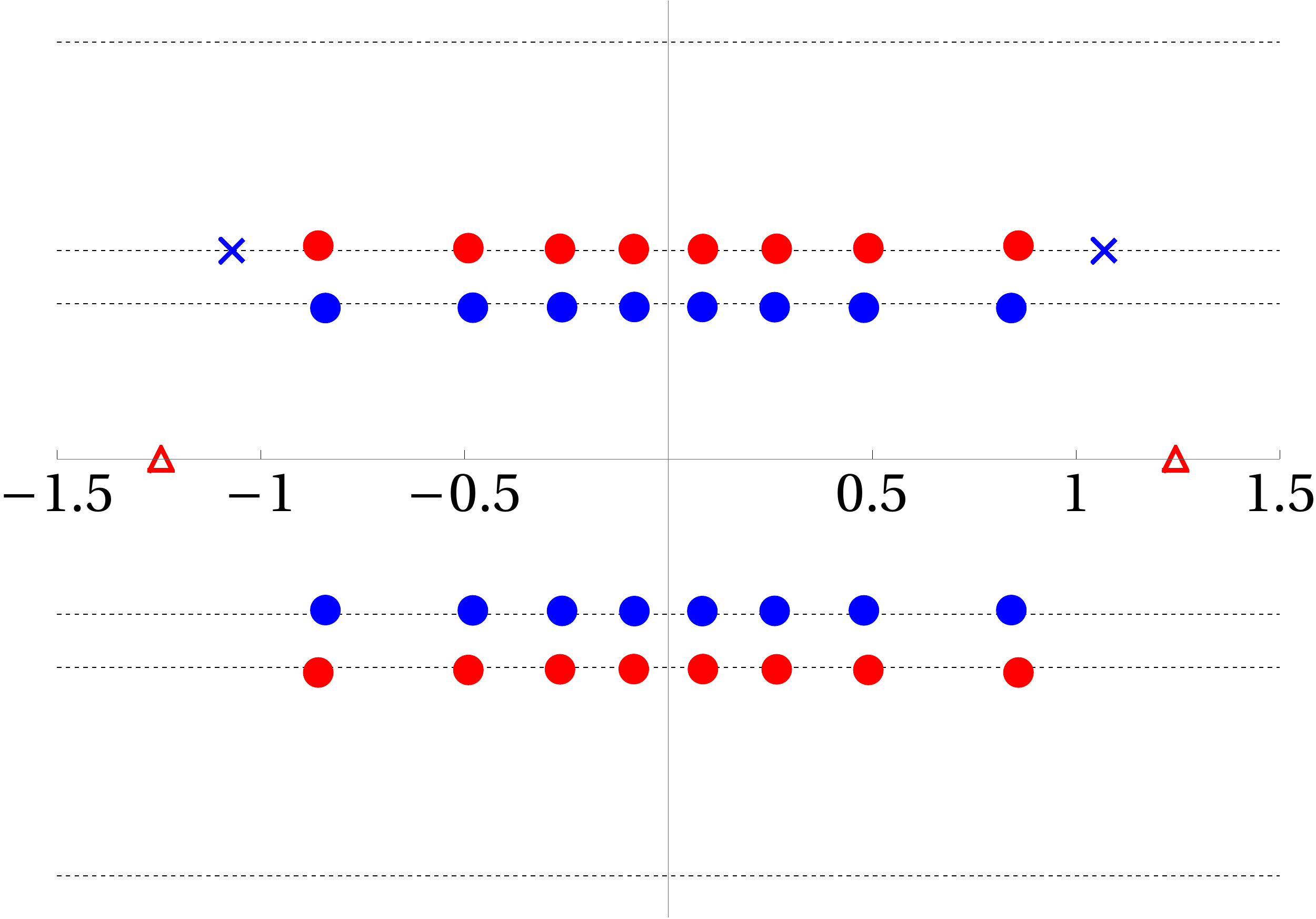}};
\node at (0.00,2.9)   {\small$\Im m(u)$};
\node[anchor=west] at  (3.75,0.00)  {\small$\Re e(u)$};
\node[anchor=west] at (3.5,2.5)   {\small$\pi$};
\node[anchor=west] at (3.5,1.35)   {\small$\frac{\pi}{2}$};
\node[anchor=west] at (3.5,0.85)   {\small$\frac{\pi}{2}-\gamma$};
\node[anchor=west] at (3.5,-0.85)   {\small$-\frac{\pi}{2}$};
\node[anchor=west] at (3.5,-1.35)  {\small$-\frac{\pi}{2}+\gamma$};
\node[anchor=west] at (3.5,-2.5)   {\small$-\pi$};

\end{tikzpicture}
}
 \end{minipage}
 \begin{minipage}[b]{.49\linewidth}
      \scalebox{0.9}{
\begin{tikzpicture}
\node at (0,0) {\includegraphics[width=0.9\linewidth]{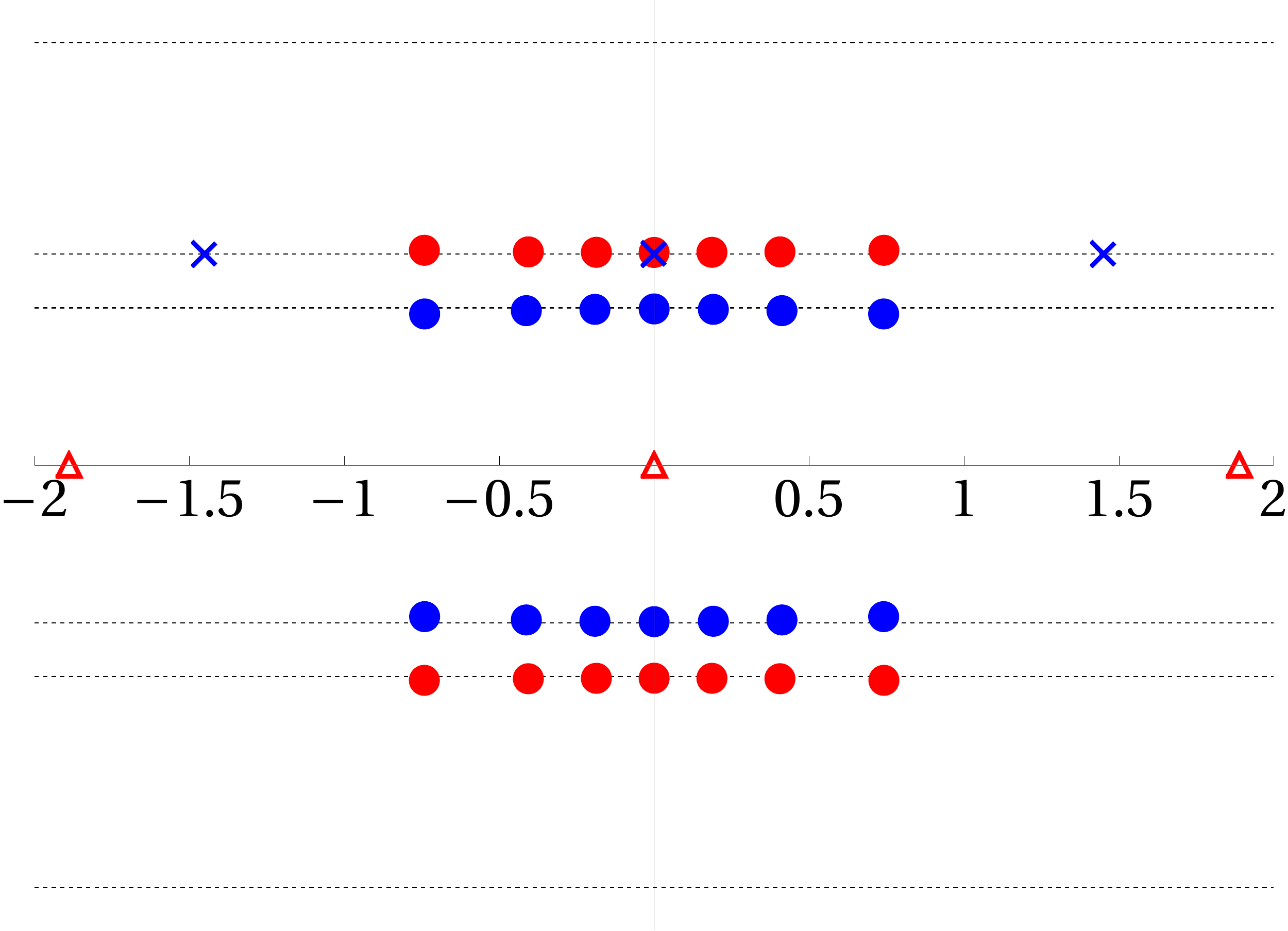}};
\node at (0.00,2.9)   {\small$\Im m(u)$};
\node[anchor=west] at  (3.75,0.00)  {\small$\Re e(u)$};
\node[anchor=west] at (3.5,2.5)   {\small$\pi$};
\node[anchor=west] at (3.5,1.35)   {\small$\frac{\pi}{2}$};
\node[anchor=west] at (3.5,0.85)   {\small$\frac{\pi}{2}-\gamma$};
\node[anchor=west] at (3.5,-0.85)   {\small$-\frac{\pi}{2}$};
\node[anchor=west] at (3.5,-1.35)  {\small$-\frac{\pi}{2}+\gamma$};
\node[anchor=west] at (3.5,-2.5)   {\small$-\pi$};
\end{tikzpicture}
}
\end{minipage}
    \caption{Left (right) plot displays the Bethe-root configuration in the complex plane of an excited state for $L=19,(18)$, $\gamma=0.4$ in the sector $ h_1=1$ and $h_2=0$. Blue (red) symbols denote level 1 (2) roots. This excitation is built by placing  2(3) level-$1$ roots ($\times$) on the line $\frac{\ri \pi }{2}$ and 2,(3) level-2 roots ($\triangle$) on the real line in addition to the bulk roots ($\bullet$). Further, one (and a half four-string) has been removed with respect to  the lowest energy state configuration in this sector. It is the state displayed in Fig.~\ref{BR_First_Conti} transformed by $u^{[2]}+\ri \pi$, and so it has the same energy.}
    \label{BR_Second_Conti}
\end{figure}
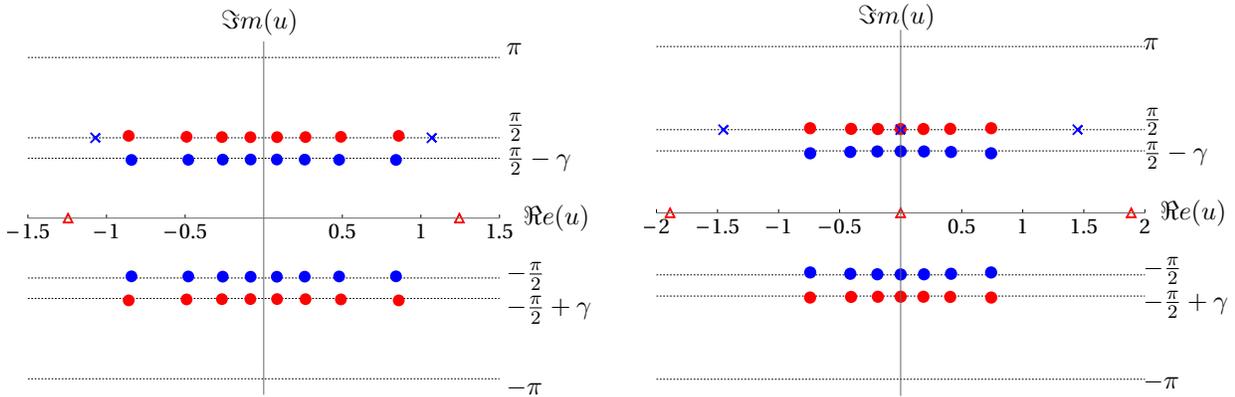

\begin{figure}[H]
    \centering
    \begin{minipage}[b]{.49\linewidth}
    \scalebox{0.9}{
\begin{tikzpicture}
\node at (0,0) {\includegraphics[width=0.9\linewidth]{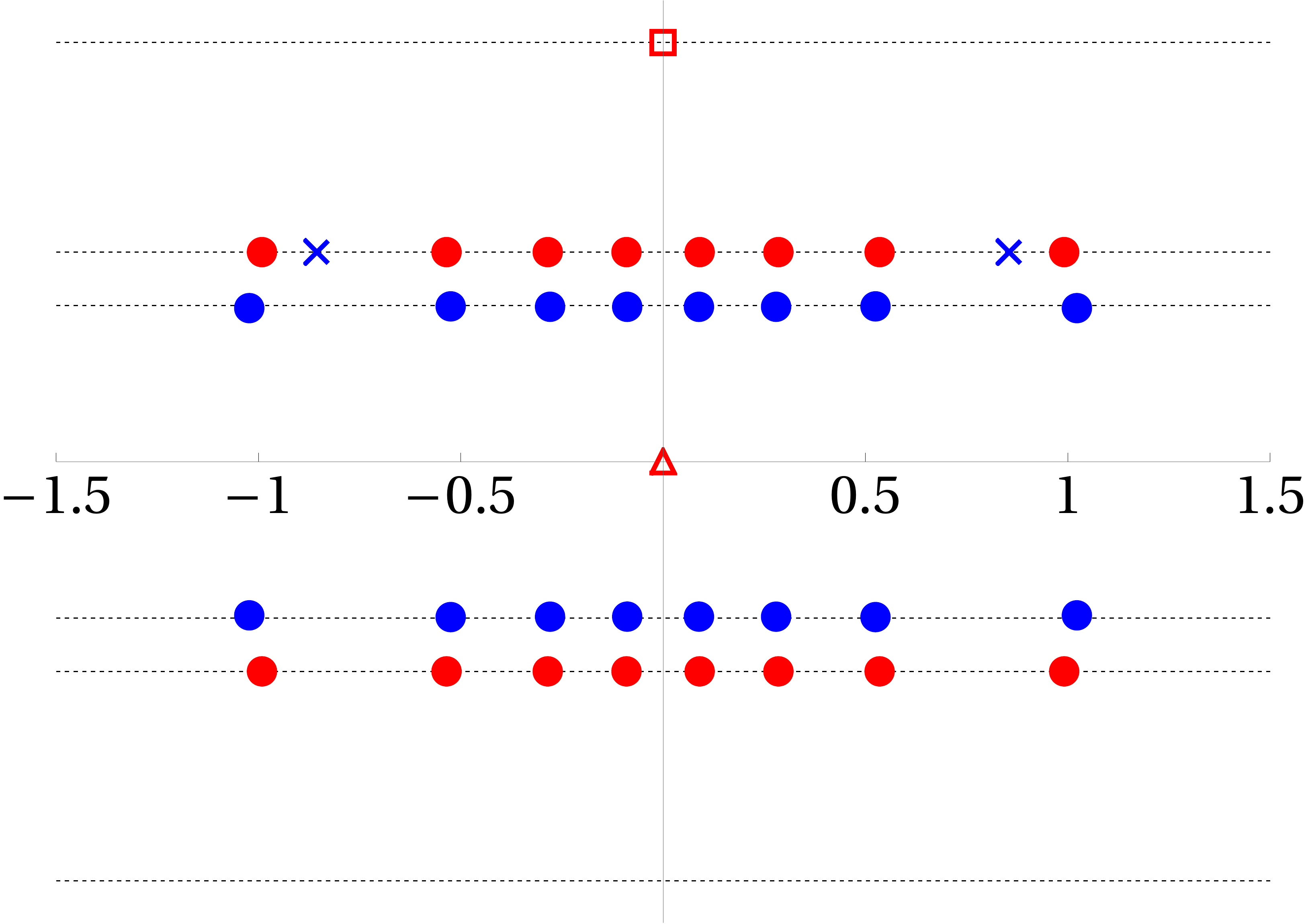}};
\node at (0.00,2.9)   {\small$\Im m(u)$};
\node[anchor=west] at  (3.75,0.00)  {\small$\Re e(u)$};
\node[anchor=west] at (3.5,2.5)   {\small$\pi$};
\node[anchor=west] at (3.5,1.35)   {\small$\frac{\pi}{2}$};
\node[anchor=west] at (3.5,0.85)   {\small$\frac{\pi}{2}-\gamma$};
\node[anchor=west] at (3.5,-0.85)   {\small$-\frac{\pi}{2}$};
\node[anchor=west] at (3.5,-1.35)  {\small$-\frac{\pi}{2}+\gamma$};
\node[anchor=west] at (3.5,-2.5)   {\small$-\pi$};

\end{tikzpicture}
}
 \end{minipage}
 \begin{minipage}[b]{.49\linewidth}
      \scalebox{0.9}{
\begin{tikzpicture}
\node at (0,0) {\includegraphics[width=0.9\linewidth]{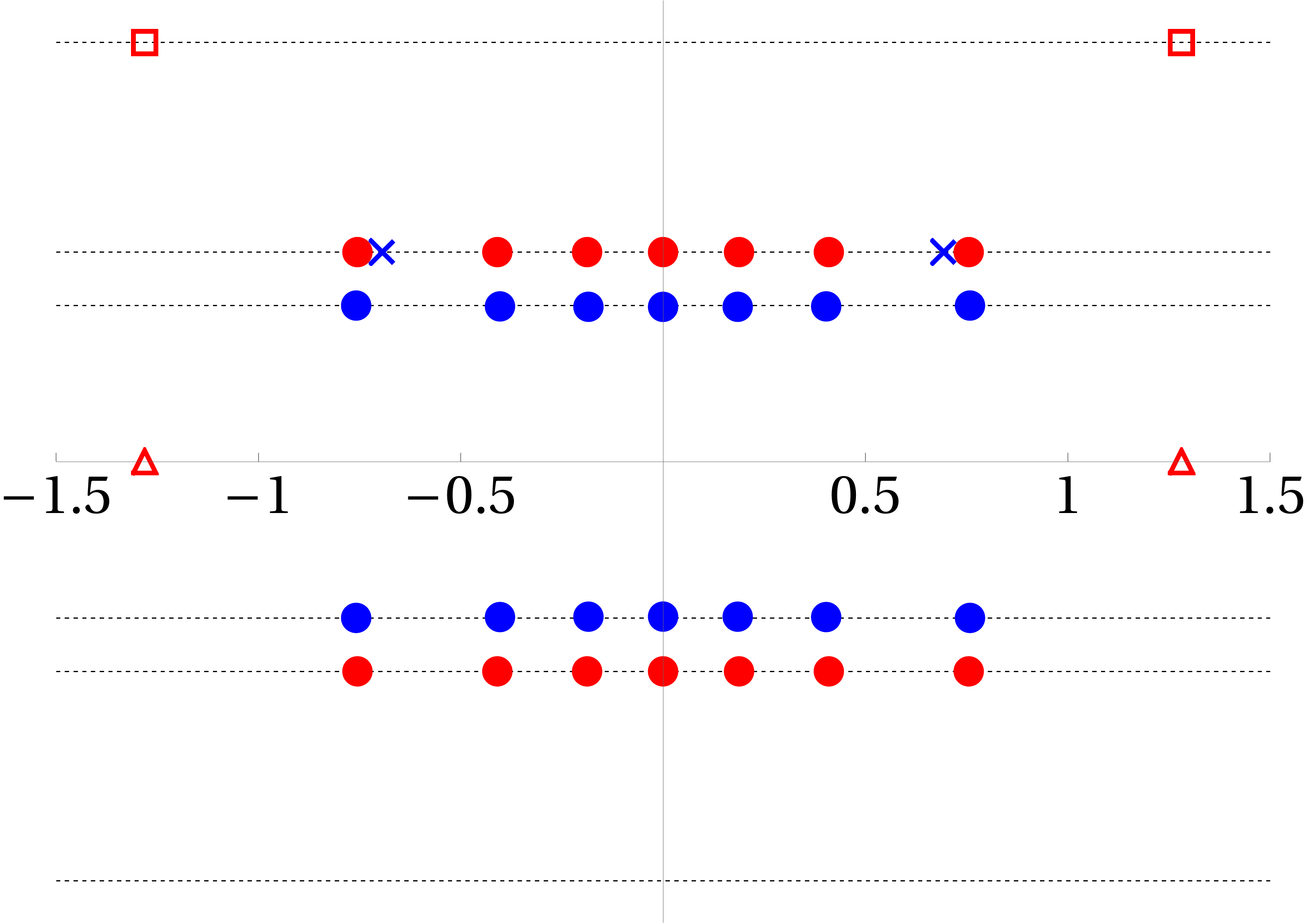}};
\node at (0.00,2.9)   {\small$\Im m(u)$};
\node[anchor=west] at  (3.75,0.00)  {\small$\Re e(u)$};
\node[anchor=west] at (3.5,2.5)   {\small$\pi$};
\node[anchor=west] at (3.5,1.35)   {\small$\frac{\pi}{2}$};
\node[anchor=west] at (3.5,0.85)   {\small$\frac{\pi}{2}-\gamma$};
\node[anchor=west] at (3.5,-0.85)   {\small$-\frac{\pi}{2}$};
\node[anchor=west] at (3.5,-1.35)  {\small$-\frac{\pi}{2}+\gamma$};
\node[anchor=west] at (3.5,-2.5)   {\small$-\pi$};
\end{tikzpicture}
}
\end{minipage}
    \caption{Left (right) plot displays the Bethe-root configuration in the complex plane of an excited state for $L=18$, $\gamma=0.4$ in the sector $h_1=0$ and $h_2=0$. Blue (red) symbols denote level 1 (2) roots. This excitation is built by placing  2(4) level-$1$ roots ($\times$) on the line $\frac{\ri \pi }{2}$ and 1,(2) level-2 roots ($\triangle$) on the real line and 1,(2) level-2 roots ($\square$) on the line $\ri \pi$ in addition to the bulk roots ($\bullet$). This is an excitation of both non-compact modes.}
    \label{BR_Mix_Contis}
\end{figure}
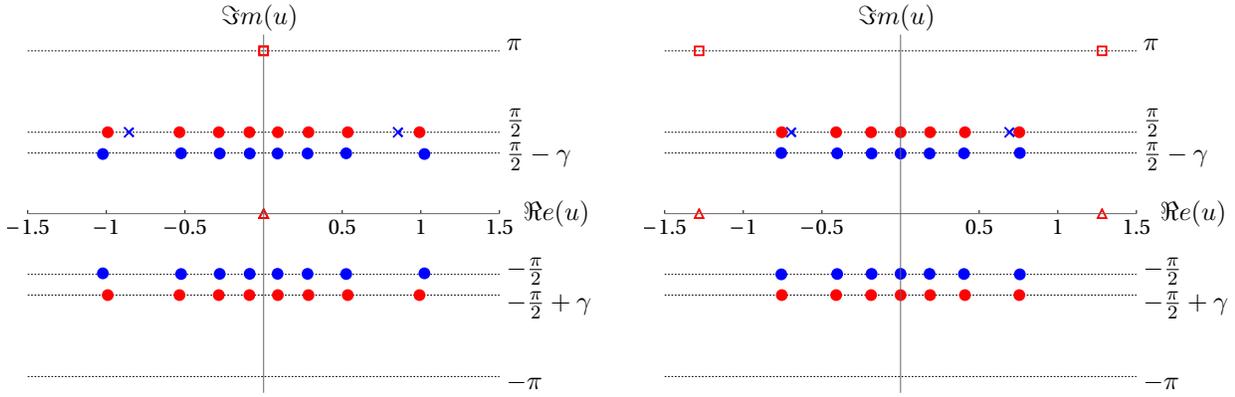

\begin{figure}[H]
    \centering
    \begin{minipage}[b]{.49\linewidth}
     \includegraphics[width=\linewidth]{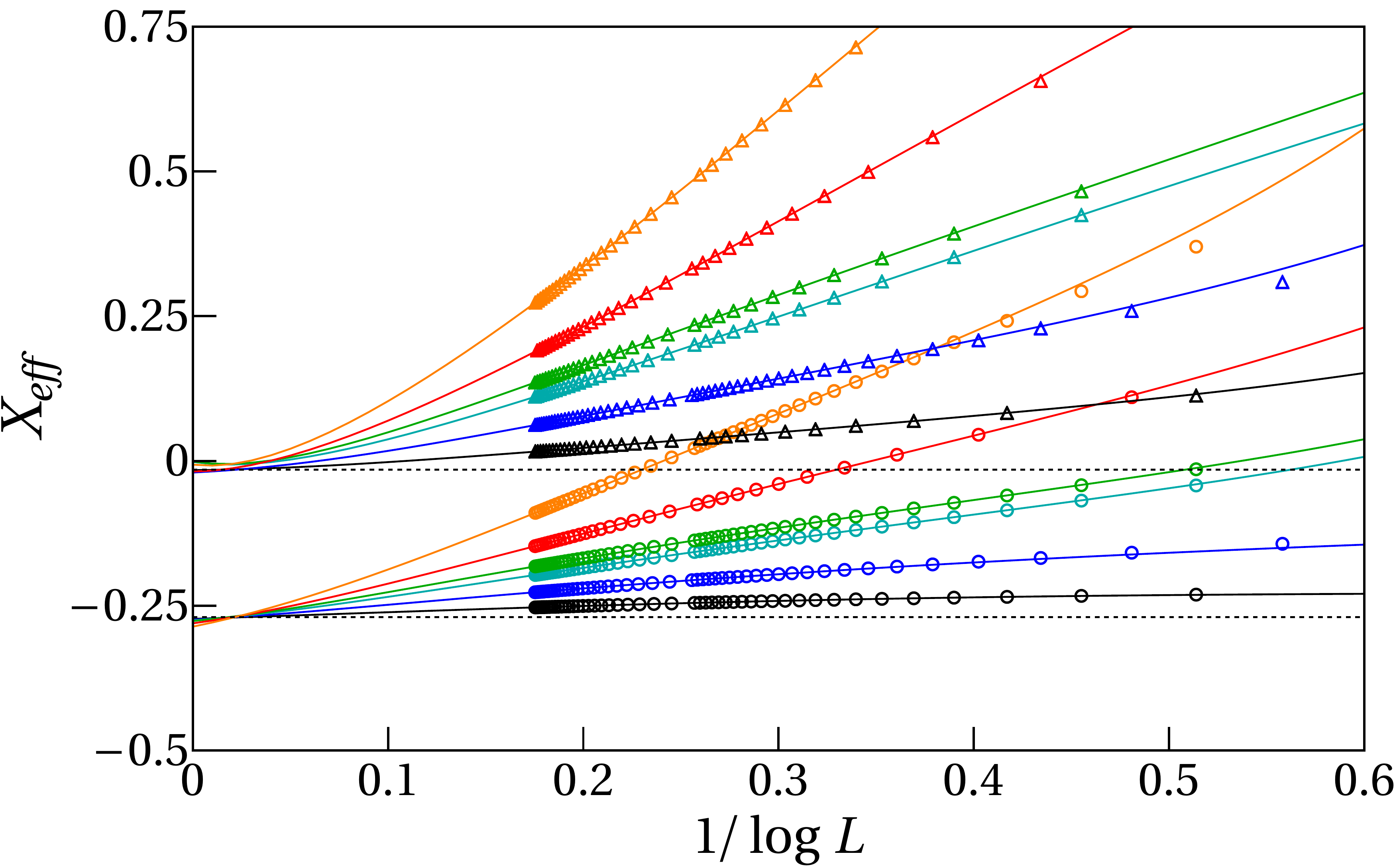}
    \end{minipage} \begin{minipage}[b]{.49\linewidth}
     \includegraphics[width=\linewidth]{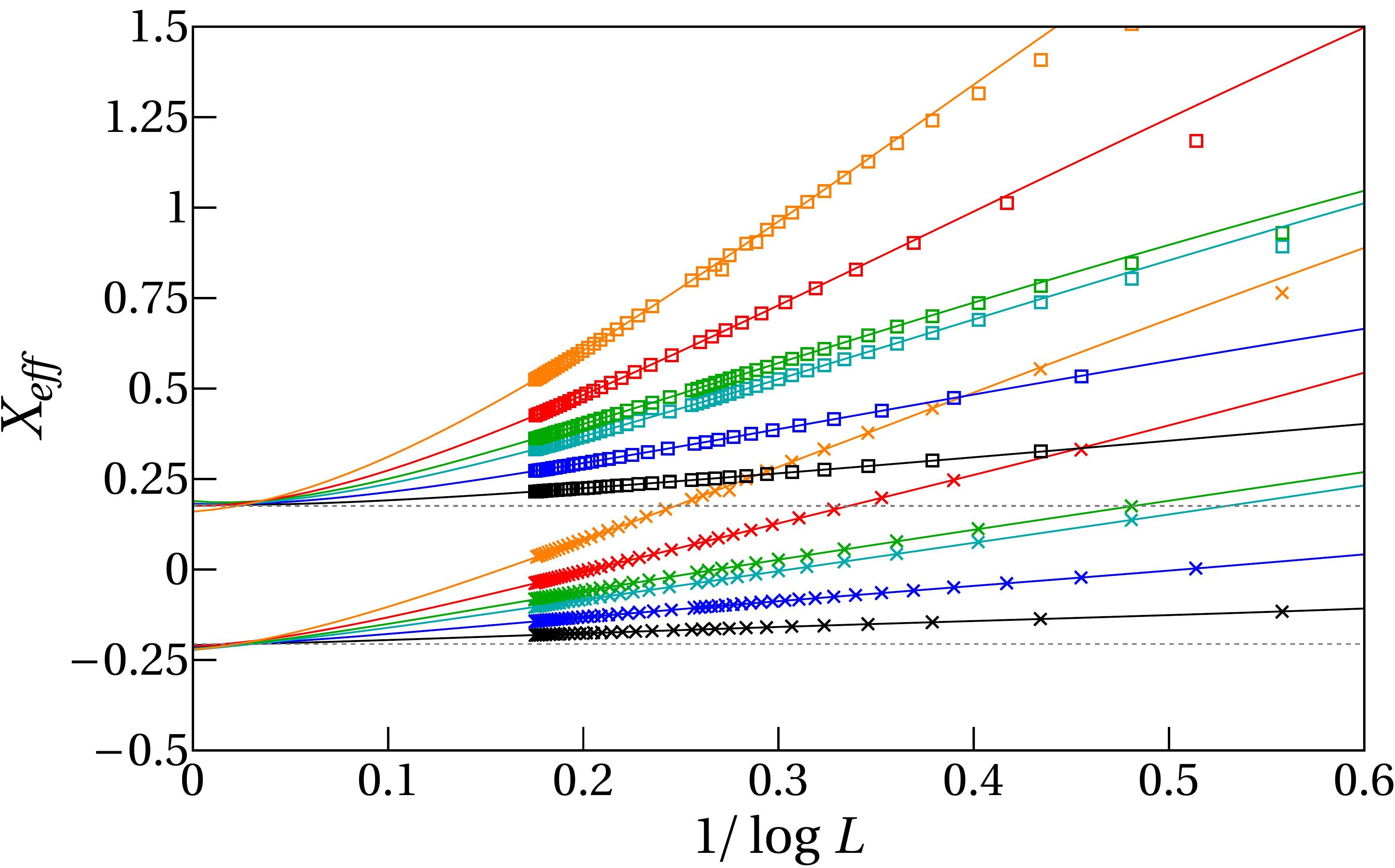}
    \end{minipage}
    \caption{ Left (right) plot displays the finite-size scaling up to $L \sim 300$ in the sectors $(h_1,h_2)$: $(1,0)$  $\circ$, $(2,1)$ $\triangle$ ($(1,1)$  $\times$, $(2,2)$  $\square$) for states with $(M_0,M_\pi)=(0,0),(0,1),(1,1),(0,2),(1,2),(2,2)$ in increasing order from below (black, blue, cyan, green, red, orange). The solid lines are obtained by a rational extrapolation. The dashed lines depict the limiting value given by (\ref{Xeff_Compact_Twist}). One can see clearly the logarithmic dependence of the scaling dimensions. The parameters are set to $\gamma=0.4$, $\phi_{1,2}=0$. }
    \label{Conti_Mixed_FS_Plot}
\end{figure}

By replacing more and more 4-strings by roots of type (\ref{i} and (\ref{ii}, one can generate an infinite tower of excitations labelled by the number $M_\pi$ of type (\ref{ii} roots. All of them flow to the same scaling dimension (\ref{Xeff_Compact_Twist}), see Fig.~\ref{Conti_1_FS_Plot}. 
Using the symmetry $u^{[2]}_j\to u^{[2]}_j+\ri \pi $ that exchanges the (\ref{ii} and (\ref{iii} types of roots, one deduces that the RG trajectories of excitations with root configuration built by mechanism (\ref{iii} instead of (\ref{ii}, see e.g. Fig.~\ref{BR_Second_Conti}, also flow to the same scaling dimensions. Let's label them by the number $M_0$ of type (\ref{iii} roots. 
Further, it turns out that combinations of the two above excitation patterns are possible, see for example Fig.~\ref{BR_Mix_Contis} for the Bethe root configuration of a mixed state of both fundamental excitations (\ref{ii} and (\ref{iii}. Further, see Fig.~\ref{Conti_Mixed_FS_Plot} to see how such states fit in within the scaling behavior of other states. 

\begin{figure}[H]
    \centering
     \includegraphics[width=0.9\linewidth]{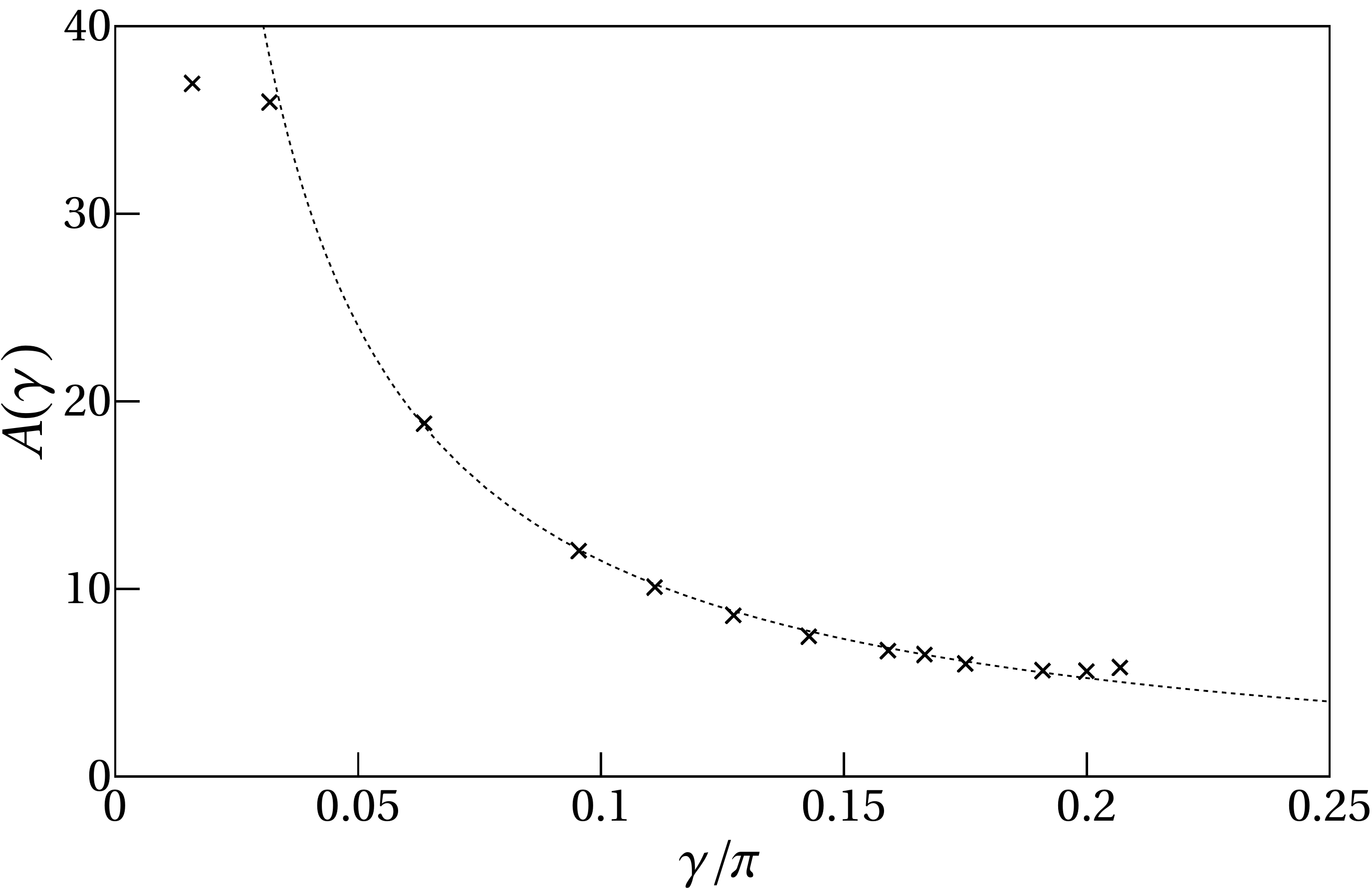}
    \caption{Amplitude $\mathcal{A}(\gamma)$ calculated via (\ref{Amplitude_Measure}) for $L^1=2000$, $L^2=1000$ for various $\gamma$-values. The dashed line is the conjecture (\ref{Amplitude}). One see a fairly good matching. At the boundaries $\gamma \approx 0, (\frac{\pi}{4})$ one sees deviations which are assumed to be due to increasing finite-size corrections, see also \cite{IkJS08}. 
    }
    \label{FS_LogAmp}
\end{figure}

The obtained numerical data for various RG trajectories with different $M_0,M_\pi$ can be used to extract the form of the logarithmic corrections: The rational extrapolation in Fig.~\ref{Conti_Mixed_FS_Plot} suggest a general quadratic decay as $\propto \frac{1}{\log (L)^2}$. Further, the existence of two excitation mechanism refines this ansatz to $\propto \frac{C_1}{\log (L)^2}+\frac{C_2}{\log (L)^2}$ with state dependent constants $C_{1,2}$. As the $Z_2$ symmetry interchanges these two contributions, we conclude that we must have $C_1=C_2$. Multiplying the numerical data with $\log (L)^2$ extrapolating $L\to \infty$, we can access, by considering ratios, the dominant state dependence for each $\ket{\Psi_L}$. This numerical work reveals the following behavior:
\begin{align}
    &X_{\rm{eff}}= X^{\rm{Com}}_{\rm{eff}}(h_1,h_2,0,0)+\frac{\mathcal{A}(\gamma)(M_0+\frac{2}{3})^2}{\log(L/L_0)^2}+     \frac{\mathcal{A}(\gamma)(M_\pi+\frac{2}{3})^2}{\log(L/L_0)^2}\,.\label{Log_Corrections}
\end{align}
Here $L_0$ is a non-universal, state dependent constant, which we do not attempt to calculate here. We are left to extract the amplitude $\mathcal{A}(\gamma)$. As it is the same for all states, we determine it by considering the effective scaling dimensions of the ground state where we expect the subleading logarithmic corrections to be the smallest:
\begin{align}
    X^{GS}_{\rm{eff}}(L)=-\frac{4}{12}+\frac{8}{9}\frac{\mathcal{A}(\gamma) }{\log(L/L_0)^2} \,.
\end{align}
Following \cite{IkJS08}, we eliminate $L_0$ by using data points for two system size $L^1$ and $L^2$:  
\begin{align}
    \mathcal{A}(\gamma)=\frac{9}{8}\Bigg[\frac{\log(\frac{L^1}{L^2})}{(X^{GS}_{\rm{eff}}(L^1)+\frac{4}{12})^{-\frac{1}{2}}-(X^{GS}_{\rm{eff}}(L^2)+\frac{4}{12})^{-\frac{1}{2}}}\Bigg]^2 \,.
    \label{Amplitude_Measure}
\end{align}
The numerical results are displayed in Fig.~\ref{FS_LogAmp}. Based on these, we conjecture that 
\begin{align}
        \mathcal{A}(\gamma)=\frac{5\pi-4\gamma}{4\gamma} \,. \label{Amplitude}
\end{align}

We want to briefly comment how the above leads to two continuous components in the spectrum of scaling dimensions. So far, we have defined the RG trajectories $\ket{\Psi_L}$ by keeping the numbers $M_{0,\pi}$ fixed, leading at first view to infinite degeneracies in the scaling limit. 
However, we can also organize the RG trajectories differently. Instead of keeping $M_{0,\pi}$ fixed, we can also let them run under the RG flow. In particular, we group states into trajectories $\ket{\Psi_L}$ such that 
\begin{align}
    M_{0}\sim \log(L)\,, \quad M_{\pi}\sim \log(L) \label{RG_M0pi}\,.
\end{align}
Here (\ref{RG_M0pi}) is subject to the constraint $M_{0,\pi}\ll L$ such that $\ket{\Psi_L}$ is still a low energy state for any finite $L$, i.e.\ its energy obeys  \eqref{Cardy_Formula}. This restriction is essential as RG trajectories leave the low energy spectrum once $M_{0,\pi}\sim L$.  In fact, it is straightforward to show within the root density approach that the state with $M_{0,\pi}=L$ is highly excited: in comparison to the energy $L e_\infty$ of the ground state (\ref{einf}), its energy is of the order $\sim L$ (similar as e.g.\ in the staggered $sl(2|1)$ superspin chain \cite{EsFS05}).  This supports the interpretation of the findings above as evidence for the existence of continuous components in the conformal spectrum labelled by two continuous quantum numbers $M_{0,\pi}/\log L$. 

The redefinition (\ref{RG_M0pi}) of RG trajectories enables that (\ref{Log_Corrections}) can tend as $L\to \infty$ to a different scaling dimension than $X^{\rm{Com}}_{\rm{eff}}(h_1,h_2,0,0)$. In fact, by suitably arranging the concrete behavior (\ref{RG_M0pi}), the scaling dimensions can take any value larger or equal to $X^{\rm{Com}}_{\rm{eff}}(h_1,h_2,0,0)$. Note that $X_{\rm{eff}}$ for trajectories with similar $M_{0,\pi}(L)$ for fixed $L$ become densely distributed ($\sim \frac{1}{\log (L)}$), leading to continuous spectrum of scaling dimensions. As the two excitation mechanisms labelled by $M_{0}$ and $M_{\pi}$ are independent of each other, we must have two continuous variables, call them $s_0\sim \frac{M_0}{\log(L)} ,s_\pi\sim\frac{M_\pi}{\log(L)}$, whose limits label the state in the scaling limit. The existence of two continua is also further supported by our finding for finite twist angles discussed below.\\

We want to stress that for the identification of the underlying CFT with two continuous components, a more rigorous definition of the scaling limit than in (\ref{RG_M0pi}) is needed. A proper scaling limit can be defined in inhomogeneous models, where the logarithmic corrections can be parameterized by a conserved operator of the lattice model, the so-called quasi-momentum operator. However, the definition of this operator in these models relies on their inhomogeneity, and is therefore not applicable to our model. For details, we refer to the extensive study of the staggered-six vertex model \cite{BKKL21a,BKKL21b} and the pioneering work  \cite{IkJS12}.

\subsubsection{Spectral flow for continuum states}

Having identified the finite-size spectrum for vanishing twist angles, we now turn to the question of what happens when these angles are tuned on. We follow the procedure described below Eq.  (\ref{Xeff_Compact_Twist}) starting from $(\phi^{ \rm{in}}_1,\phi^{ \rm{in}}_{2})=(0,0)$ and iterating to the higher twists. This procedure can be technically involved, as certain roots of a given configuration can tend to infinity as the twists approach certain values. If these specific twist values are exceeded, then the infinite roots come back to a finite value. To avoid numerical problems caused by these infinities, it is suitable to transform to a different coordinate set. By using $\zeta=\re^{-u}$ infinitely large roots are mapped to zero in this coordinate frame. 

\begin{figure}[H]
    \centering
    \begin{minipage}[b]{.49\linewidth}
     \includegraphics[width=\linewidth]{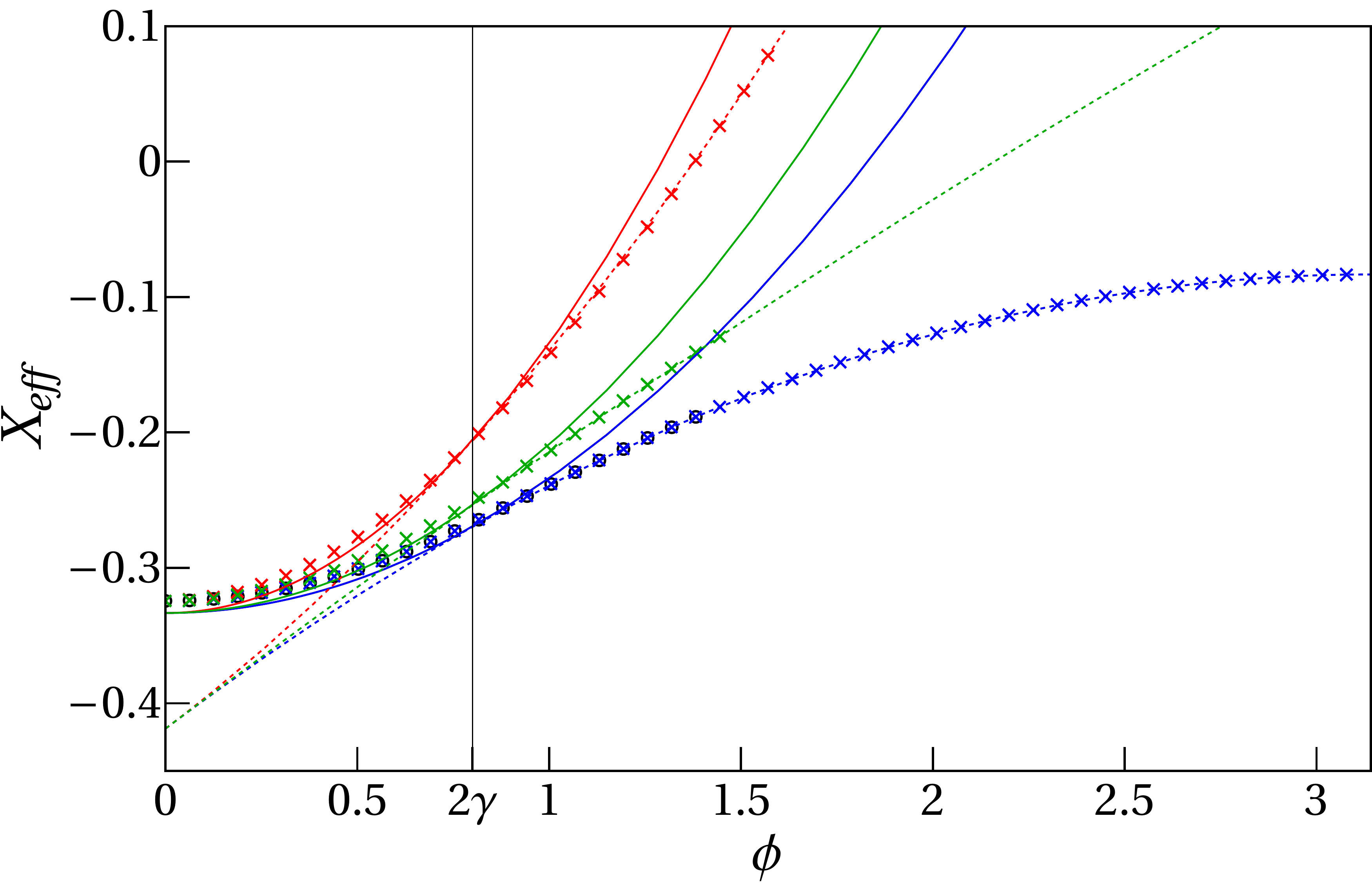}
    \end{minipage} \begin{minipage}[b]{.49\linewidth}
     \includegraphics[width=\linewidth]{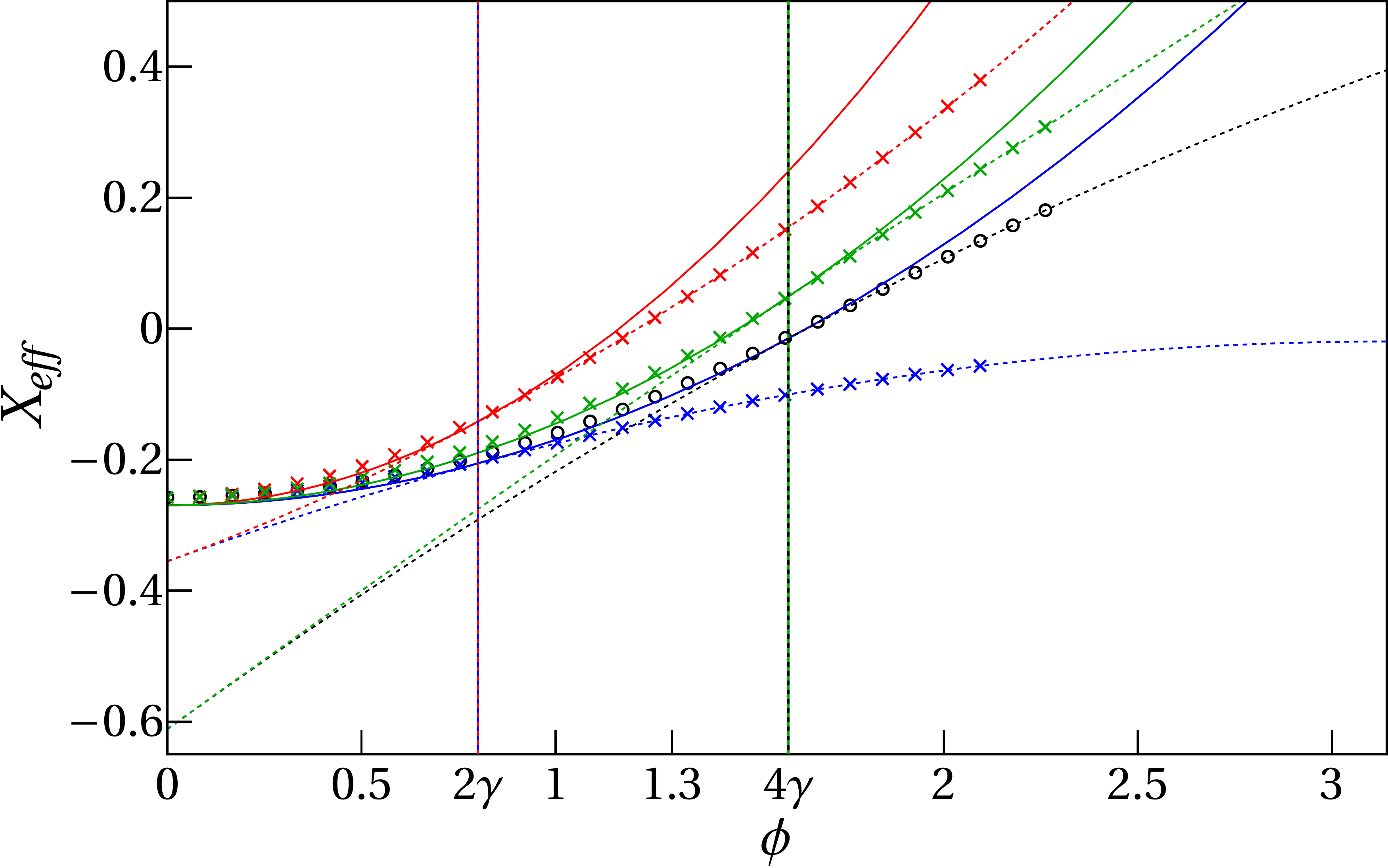}
    \end{minipage}
    \caption{The left (right) plot displays the effective scaling dimensions for the lowest state in the continuum in the sectors $(h_1,h_2)=(0,0)$, $((1,0))$ for  $L=2000$, $(1999)$, under various twists $(\phi_1,\phi_2)=(\phi,0),(0,\phi),(\phi,\phi),(\phi,\frac{1}{2}\phi)$ (Black, Blue, Red, Green). The solid lines display the expected behavior (\ref{Xeff_Compact_Twist}) for small twist angles excluding the strong logarithmic corrections. The crosses or circles display the numerical data obtain from the Bethe ansatz for twist angles as far as possible in the numerical procedure. The vertical lines designate the critical twist values, where the agreement with (\ref{Xeff_Compact_Twist}) breaks down.  The dashed lines indicate the conjectured formula (\ref{Discrete_Xeff}) for the scaling dimensions valid beyond the critical points. Note that the matching with the conjecture is extremely accurate. We interpret this as the emergence of discrete states having less logarithmic corrections.}
    \label{twisted_Xeff}
\end{figure}

We start by discussing the lowest states ($M_0=M_\pi=0$) in the two continua first. 
Some of our results are represented in Fig.~\ref{twisted_Xeff}. One can see that for small twist angles, the scaling dimensions follow  (\ref{Log_Corrections}) but with the first term replaced by (\ref{Xeff_Compact_Twist}) with non-vanishing twist. However, after critical values of the twists given by 

\begin{align}
    \phi^c_{1,2}=2\gamma(h_{1,2}+1)=\frac{2 \pi  (h_{1,2}+1)}{\kk} \,,
\end{align}
the behavior changes drastically to 
\begin{align}
        X_{\rm{eff}}= X^{\rm{Com}}_{\rm{eff}}(h_1,h_2,\phi_1,\phi_2)-\frac{(h_{1,2}+1-\kk \frac{\phi_{1,2}}{2\pi})^2}{2( \kk-2)} \,, \qquad \phi^{c}_{1,2}<\phi_{1,2}<\widetilde{\phi}^{c}_{1,2}\,, \label{Discrete_Xeff}
\end{align}
with the absence of logarithmic corrections. Numerical work suggests that the choice between $1$ and $2$ in the above formulae seems to taken in a way such that the second term in (\ref{Discrete_Xeff}) always incorporates the bigger twist angle and, if both twists are equal, minimizing the critical twist angle (see Fig.~\ref{twisted_Xeff}).  It turns out that (\ref{Discrete_Xeff}) is valid just until the twist exceeds another critical twist angle $\widetilde{\phi}^{c}_{1,2}$. For example, for twisting only with $\phi_1$ or $\phi_2$ in the lowest sector $h_1=h_2=0$, we find that
\begin{align}
   \left.\widetilde{\phi}^{c}_{1,2}\right|_{h_1=h_2=0}=2\pi-\left.\phi^{c}_{1,2}\right|_{h_1=h_2=0} \,.
\end{align}
\color{black}
So far, we have considered only the lowest states in the continua. If we twist excited states, their scaling dimensions follow (\ref{Discrete_Xeff}) but again spoiled by decreasing logarithmic corrections. To further investigate this phenomenon, we have searched for twists for which the Bethe root configurations are again regular enough to define RG trajectories. We find that a suitable point is $(\phi_1,\phi_2)=(\pi,0)$. Here, the Bethe roots parameterizing the low-energy states consists mainly of
\begin{equation}
\begin{aligned}
  \BR{1}{} \longrightarrow&\quad x_j+ \frac{\ri\pi}{2}-\ri\gamma-\ri\epsilon^{[1]}_j\,, \quad x_j- \frac{\ri\pi}{2}+\ri\gamma+\ri\epsilon^{[1]}_j\, \\
  \BR{2}{} \longrightarrow&  \quad z_l+\frac{\ri \pi}{2}\,\,\,, \quad l=1,\dots,M_z, \\
  & \quad w_k-\frac{\ri \pi}{2}, \quad  k=1,\dots,M_w  \,.
\end{aligned}
\end{equation}
In addition there are level-1 roots sitting exponentially close to the following values depending on the parity of $dN :=M_z-M_w$:
\begin{equation}
\label{Singular_roots}
\begin{aligned}
    \ri \gamma,-\ri \gamma \quad &\text{if } dN  \text{ even}\\
    2\ri \gamma,0, -2\ri \gamma \quad &\text{if } dN  \text{ odd} 
\end{aligned}
\end{equation}
Plots of typical configurations are shown in Fig.~\ref{BR_twisted}. It turns out that states with different distributions $dN$ of level-2 roots on the two lines $\pm \frac{\ri \pi}{2}$ flow to the same conformal dimensions (\ref{Discrete_Xeff}) with $\phi_1=\pi$ and $\phi_2=0$, see e.g. Fig.~\ref{Xeff_twist_pi}. We find that the scaling of dimensions of excited states for $\phi_1=\pi, \phi_2=0$ are given by 
\begin{align}
    X_{\rm{eff}}=\left.X^{\rm{Com}}_{\rm{eff}}(h_1,h_2,\phi_1,\phi_2)\right|_{\phi_1=\pi, \phi_2=0}-\left.\frac{(h_{1}+1-\kk \frac{\phi_1}{2\pi})^2}{2( \kk-2)}\right|_{\phi_1=\pi}+\Tilde{\mathcal{A}}(\gamma)\frac{dN^2}{\log(L/\Tilde{L}_0)^2}\,.\label{Xeff_pi}
\end{align}
where again $\Tilde{L}_0$ is a non-universal constant which we do not attempt to calculate here. Further, the above formula also holds true for small deviations around the twist angles i.e. $\phi_1\approx \pi$, $\phi_2\approx 0$ with the obvious modifications. 
We conjecture that the amplitude $\Tilde{\mathcal{A}}(\gamma)$ is given by 
\begin{align}
    \Tilde{\mathcal{A}}(\gamma)=\frac{2 (2-5 \gamma ) \gamma }{3 (1-4 \gamma )^2}\,.
\end{align}
In order to interpret these results, let us recall that the scaling limit of models such as the staggered-six vertex model or the $A^{(2)}_2$-model possess one continuous parameter, call it $s_*$. Besides this class of states, there also exist states where $s_*$ takes values in a discrete set. The lattice regularization of those states does not possess logarithmic corrections. Further, a state belonging to the family with continuous $s_*$ can become a discrete state under a twist \cite{BKKL21a,VeJS14}. We interpret the above finding in an analogous way for our model. Consider a state whose scaling limit is described by the two continuous variables $s_0,s_\pi$. Under a twist, one of the continuous variables changes its class to the discrete one, while the other remains in the continuous family. The latter still induces logarithmic corrections on the level of the lattice regularization as seen in (\ref{Xeff_pi}). \\
\begin{figure}[H]
    \centering
    \begin{minipage}[b]{.49\linewidth}
    \scalebox{0.9}{
\begin{tikzpicture}
\node at (0,0) {\includegraphics[width=0.9\linewidth]{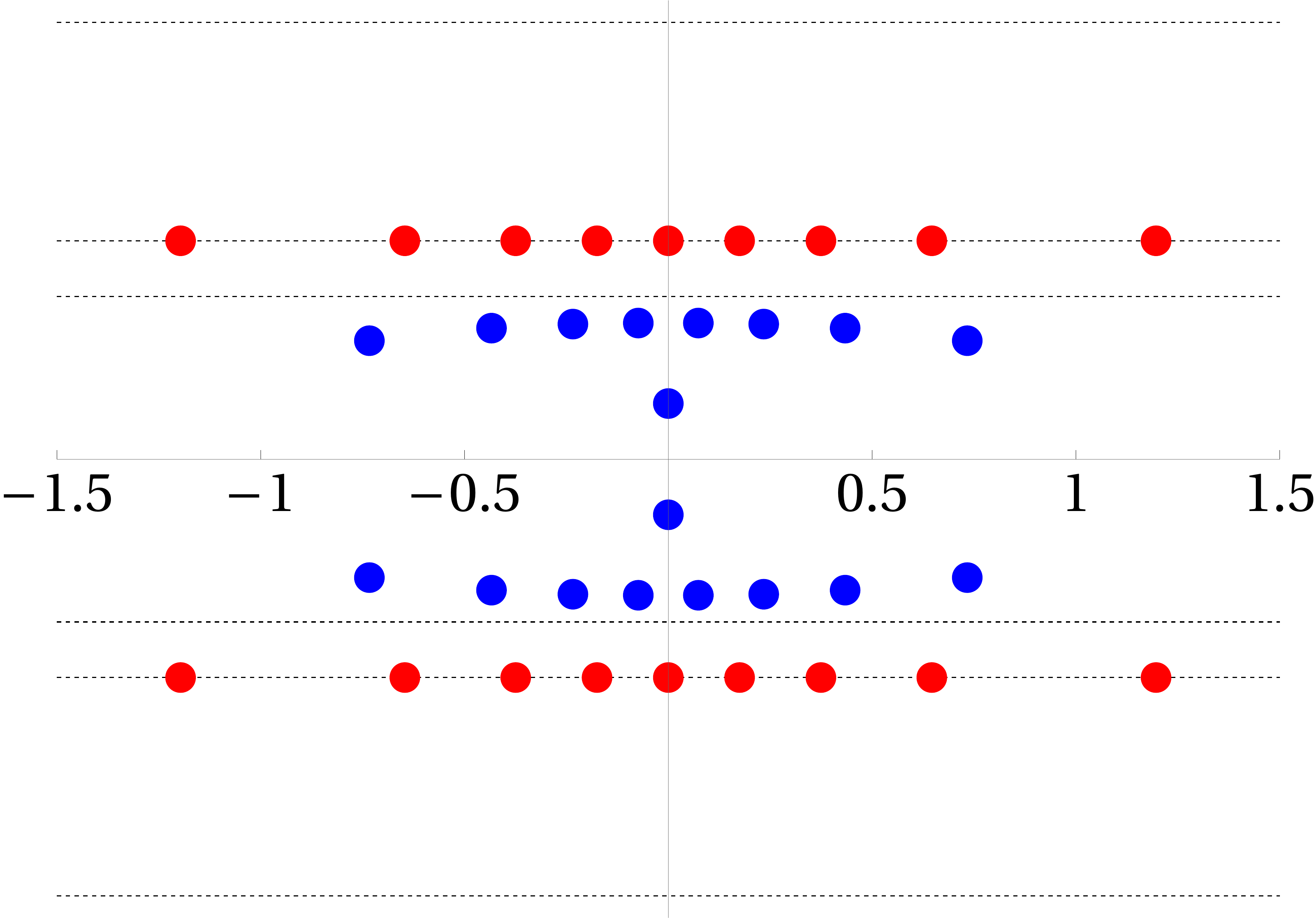}};
\node at (0.00,2.9)   {\small$\Im m(u)$};
\node[anchor=west] at  (3.75,0.00)  {\small$\Re e(u)$};
\node[anchor=west] at (3.5,2.5)   {\small$\pi$};
\node[anchor=west] at (3.5,1.35)   {\small$\frac{\pi}{2}$};
\node[anchor=west] at (3.5,0.85)   {\small$\frac{\pi}{2}-\gamma$};
\node[anchor=west] at (3.5,-0.85)   {\small$-\frac{\pi}{2}$};
\node[anchor=west] at (3.5,-1.35)  {\small$-\frac{\pi}{2}+\gamma$};
\node[anchor=west] at (3.5,-2.5)   {\small$-\pi$};

\end{tikzpicture}
}
 \end{minipage}
 \begin{minipage}[b]{.49\linewidth}
      \scalebox{0.9}{
\begin{tikzpicture}
\node at (0,0) {\includegraphics[width=0.9\linewidth]{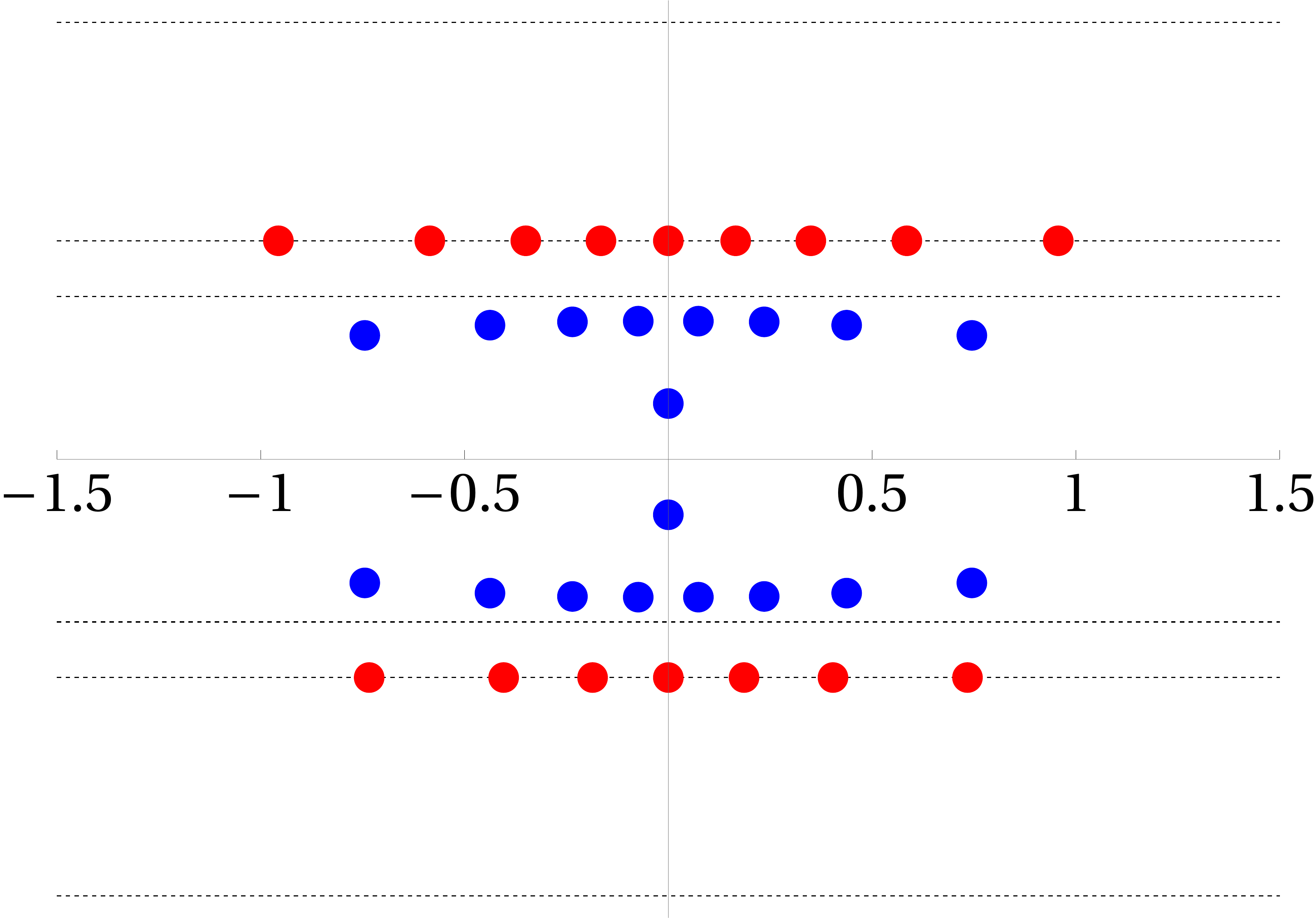}};
\node at (0.00,2.9)   {\small$\Im m(u)$};
\node[anchor=west] at  (3.75,0.00)  {\small$\Re e(u)$};
\node[anchor=west] at (3.5,2.5)   {\small$\pi$};
\node[anchor=west] at (3.5,1.35)   {\small$\frac{\pi}{2}$};
\node[anchor=west] at (3.5,0.85)   {\small$\frac{\pi}{2}-\gamma$};
\node[anchor=west] at (3.5,-0.85)   {\small$-\frac{\pi}{2}$};
\node[anchor=west] at (3.5,-1.35)  {\small$-\frac{\pi}{2}+\gamma$};
\node[anchor=west] at (3.5,-2.5)   {\small$-\pi$};
\end{tikzpicture}
}
\end{minipage}
    \caption{Left (right) plot displays the Bethe-root configuration in the complex $u$-plane of an excited state for $L=18$, $\gamma=0.4$ in the sector $h_1=0$ and $h_2=0$. Blue (red) symbols denote level 1 (2) roots. The left figure shows the ground state configuration, while the right 
    excitation is built by unbalancing $dN=4$ the number of level-2 on the lines $\pm\frac{\ri \pi}{2}$. The two level-1 roots with vanishing real part have imaginary close to $\pm \gamma$. }
    \label{BR_twisted}
\end{figure}
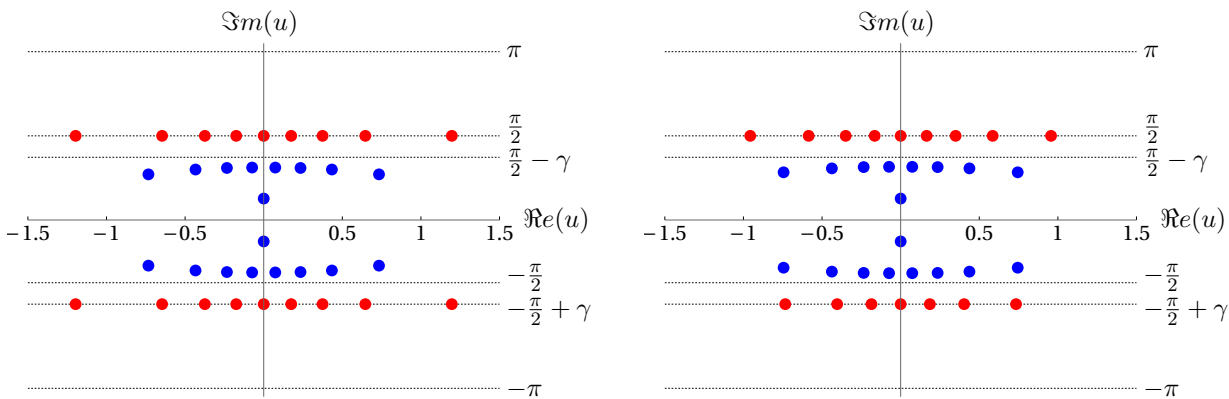
\begin{figure}[H]
    \centering
    \includegraphics[width=0.9\textwidth]{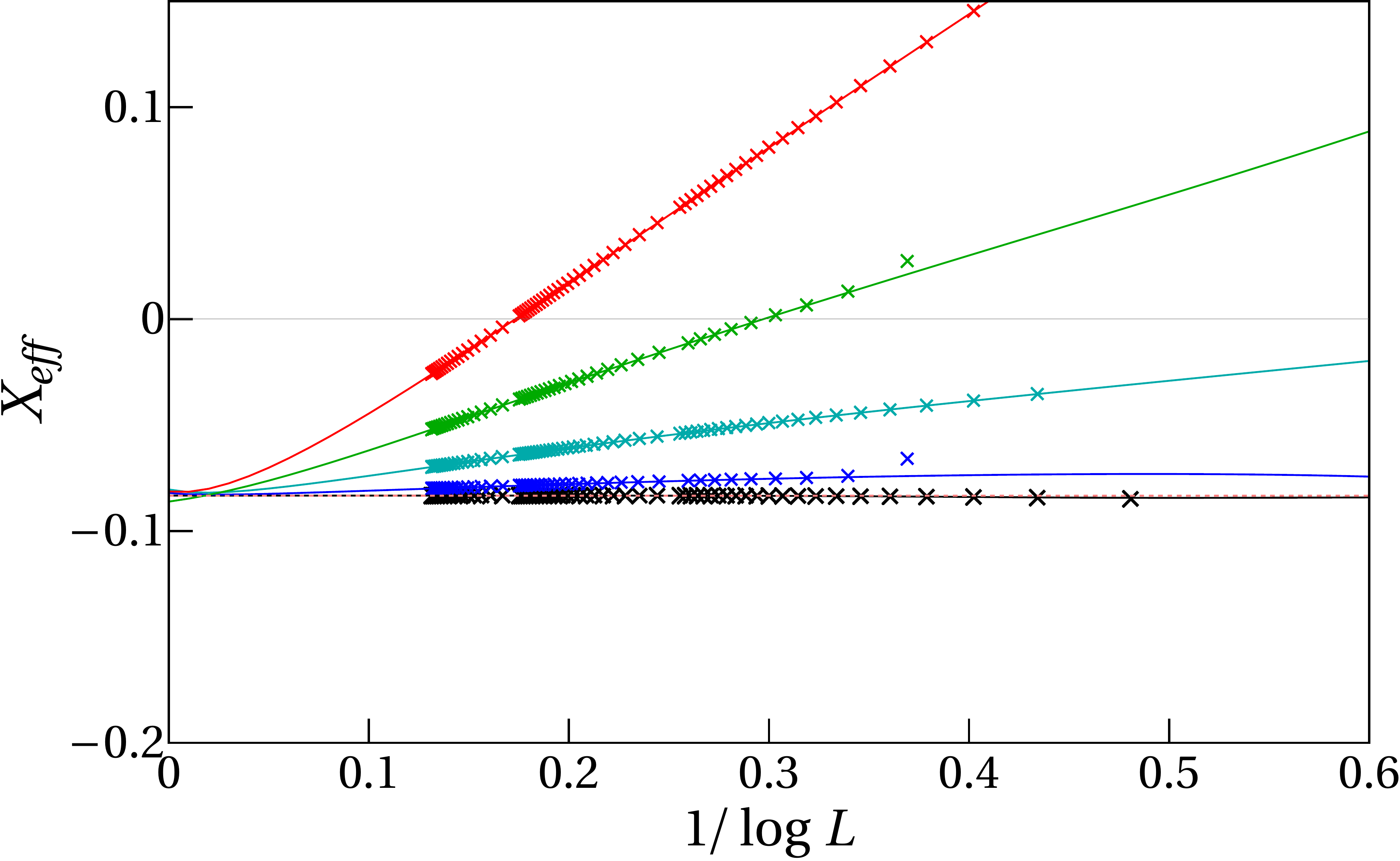}
    \caption{Effective scaling dimensions up to $L\sim 2000$ for $\phi_1=\pi$, $\phi_2=0$, $\gamma=0.4$ in the sector $h_1=h_2=0$ for $dN=0,1,2,3,4$ in increasing order from below (black, blue, cyan, green, red). The crosses are the numerical data obtain from the Bethe ansatz. The solid lines are rational extrapolation. Further, the dashed pink line is given by the constant limit value (\ref{Discrete_Xeff}). In order to obtain the numerical data, we have assumed that the roots which are exponentially close to (\ref{Singular_roots}) actually sit on these values. This leads to a small offset (see the green and blue crosses on the far right) for small system sizes, as here the approximation is inducing an error. }
    \label{Xeff_twist_pi}
\end{figure}
Ultimately, this conjecture should imply the existence of purely discrete states (apart from the $dN=0$ state)  without any logarithmic corrections. We have checked that this is indeed the case. Starting from the twist $(\phi_1,\phi_2)=(\pi,0)$, we turn on the second twist significantly. We find that the first excited states $dN=1,2$ in the sector $h_1=0=h_2$ become purely discrete states when the second twist angle exceeds the critical value $\phi_2=2\gamma$. It has effective scaling dimensions
\begin{align}
    X_{\rm{eff}}=X^{\rm{Com}}_{\rm{eff}}(h_1,h_2,\pi,\phi_2)-\frac{(h_{1}+1-\kk \frac{\pi}{2\pi})^2}{2( \kk-2)}-\frac{(h_{2}+1-\kk \frac{\phi_{2}}{2\pi})^2}{2( \kk-2)} \quad \text{with} \quad \phi_2>2\gamma \,. \label{Pure_Dis}
\end{align}

\begin{figure}[H]
    \centering
    \includegraphics[width=0.9\textwidth]{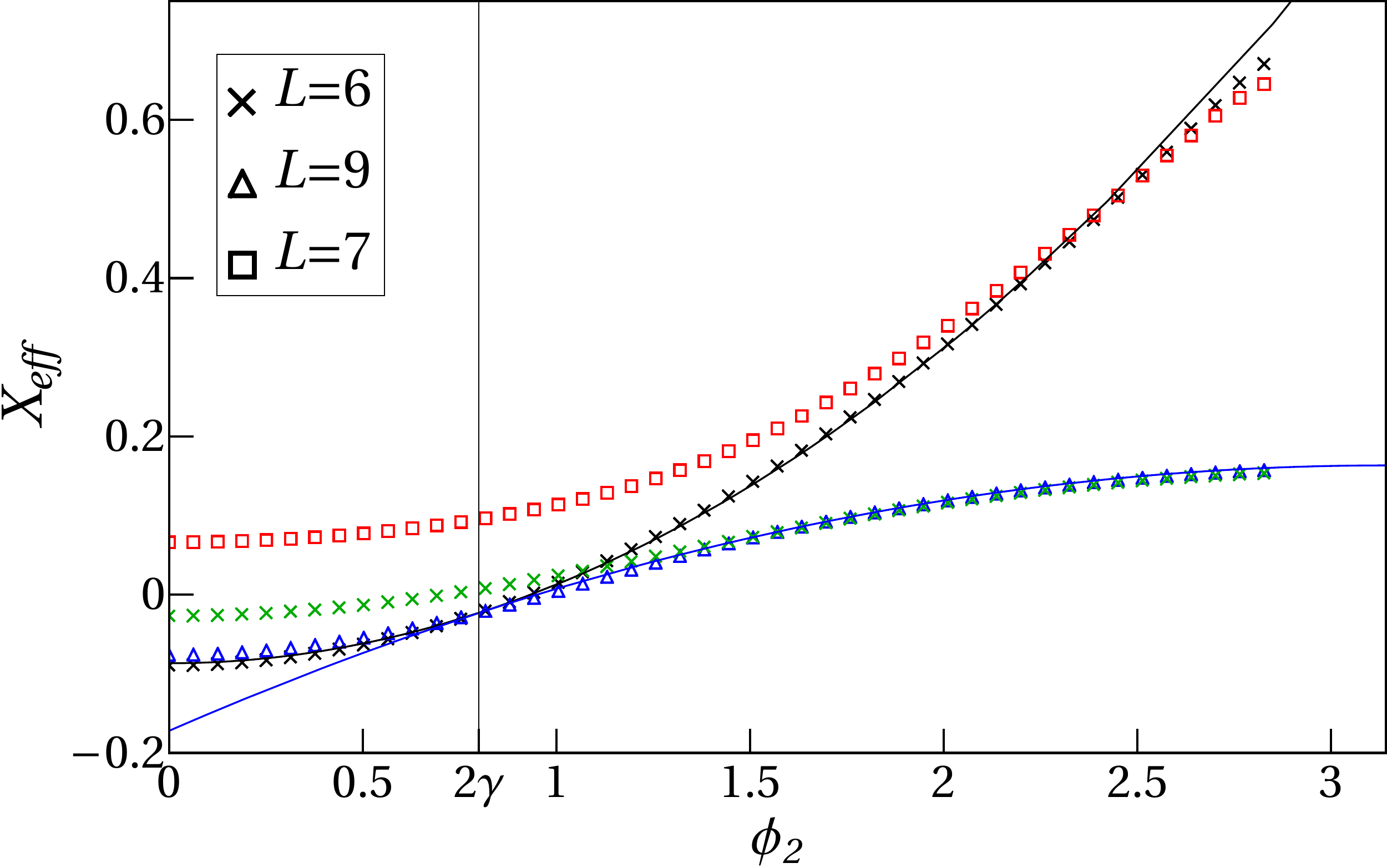}
    \caption{Effective scaling dimensions for small $L$ and $\phi_1=\frac{9\pi}{10}$, $\gamma=0.4+\ri \epsilon$ in the sector $h_1=h_2=0$ for $dN=0,1,2,3$ (black, blue, green, red) under variation of the second twist $\phi_2$. The sightly complex value of the anisotropy and offset of the first twist angle from $\pi$ is due to numerical purposes. The black solid line displays $X^{\rm{Com}}_{\rm{eff}}(h_1,h_2,\frac{9\pi}{10},\phi_2)$ while the blue line is given by (\ref{Pure_Dis}). The black vertical line hallmarks the appearance of a purely discrete state.  }
    \label{Xeff_twist_pi_phi_2}
\end{figure}

Note that this check can be done on the level of small $L$, as we expect that this state does not possess any logarithmic corrections, see Fig.~\ref{Xeff_twist_pi_phi_2}. The purely discrete  scaling dimensions (\ref{Pure_Dis}) are valid for large twist angle only; however, one can analytically continue back the scaling dimensions (\ref{Pure_Dis}) to zero twist. For the lowest state with $h_1=h_2=0$, we obtain in this way:
\begin{align}
    X_{\rm{eff}}=-\frac{4}{12}+\frac{1}{\kk-2} \,.
\end{align}
Assuming that the conformal weights $h,\bar{h}$ vanish in this procedure as it is for example in the staggered six-vertex model or the $A^{(2)}_2$ model \cite{VeJS14}, we obtain on speculative grounds that the central charge is given by 
\begin{align}
    c&=4-\frac{12}{\kk-2}=2\left(2-\frac{6}{\kk-2} \right) \,,
    \label{true_C}
\end{align}
which formally coincides with two black hole CFTs \cite{MaHi01,HaPT02,Scho06}. 
\section{Conclusion}
Starting from the periodic $D^{(2)}_3$ spin chain \cite{Resh87}, we have generalized it to the quasi-periodic case. The boundary conditions are found to be parameterized by two twist angles $\phi_1$, $\phi_2$. The appearance of the twists $\phi_1$, $\phi_2$ can be accounted for in the analytic Bethe ansatz such that the model can be exactly solved in the Bethe ansatz sense \eqref{TQtwist}, \eqref{Twisted_BAE}. As the rank of $D^{(2)}_3$ is two, the Bethe ansatz is two-level nested. As usual, the model possesses an infinite family of commuting operators \eqref{TT}. Among these, a local Hamiltonian can be defined in the standard way by the first logarithmic derivative of the transfer matrix \eqref{eq:Hamil}. 

The model has a rich symmetry structure, see \eqref{Z2symmetry}, \eqref{W_Sym2}, \eqref{CPT}. The most interesting result is that a generalization \eqref{Z2symmetry} of the $Z_2$-symmetry of the lower-rank case $D^{(2)}_2$ is identified. It is related to the quasi-periodicity of the R-matrix \eqref{quasiperiodicity}; and on the level of the Bethe ansatz, this symmetry maps states among each other whose level-$2$ Bethe roots differ by $\ri \pi$. Further, we have found that the transfer matrix is CPT-invariant \eqref{CPT}, while on the Hamiltonian level this symmetry reduces to  CP-symmetry \eqref{CPHam}.

Turning to the analysis of the scaling limit, we have concentrated on the regime of the anisotropy $\gamma\in (0,\frac{\pi}{4})$. The spin chain is found to be critical, as it possesses gapless excitations with a linear dispersion relation. Hence, the effective theory of its low-lying excitations arising in the thermodynamic limit $L\to \infty$ should be governed by a conformal field theory. We have identified certain classes of the low-lying energy states which are parameterized by the $U(1)$-charges. We have found that their effective scaling dimensions give rise to two compact modes in the scaling limit. More precisely, these modes mimic two compact bosons with zero winding \eqref{Xeff_Compact_Twist}. Indeed, despite considerable numerical effort, we did not find any non-zero winding states. Whether non-trivial winding states exist is left open for future investigation. 

In addition to the two compact modes, we found two types of decreasing logarithmic corrections. The corrections are generated by the number of level-2 Bethe roots on the real line and on the line with imaginary part $\pi$.  We provide evidence that these logarithmic corrections give rise to two non-compact degrees of freedom in the scaling limit. The two non-compact modes are interchanged by the $Z_2$ symmetry \eqref{Z2symmetry}. Furthermore, we have considered the influence of large twists. We found that, beyond certain critical twist angles, some of the logarithmic corrections disappear. We interpret this phenomenon as the emergence of discrete states under twists. For the case of $\phi_1=\pi$ and small $\phi_2$, we find that one of the continua becomes totally discrete, while the other persists.  For the extreme case of $\phi_1=\pi$ and large $\phi_2$, we observe the existence of purely discrete states \eqref{Pure_Dis}. By analytical continuation of its effective scaling dimension to zero twists, and under the assumption that the conformal dimensions vanish there \cite{VeJS14,FrHo17}, we access the true central charge \eqref{true_C}. Formally, it agrees with the sum of two Black Hole CFT central charges. One copy of this CFT describes the scaling limit of the lower-rank $D^{(2)}_2$ model \cite{NeRe21a,IkJS12,BKKL21a}. 

 For a rigorous identification of the underlying CFT, we would need a conserved operator that parameterizes the non-compact degrees of freedom on the lattice.  Such an operator, the so-called quasi-momentum, has so far been defined only in staggered models in which either the representation of the R-matrix \cite{FrHo17} or the associated spectral parameter \cite{IkJS12,KoLu23} of the quantum spaces varies periodically along the chain. Its definition relies on the type of inhomogeneity; hence, this construction is not applicable for the case of the homogeneous $D^{(2)}_3$ model we are considering here. The search for such an operator might be an interesting research direction, supporting the analysis of the scaling limit. It could help with the identification of the space of states in the scaling limit, and especially the calculation of the density of states of the continua. 
 
It should be possible to study the influence of open boundary conditions for selecting certain sectors of the underlying CFT, as has been done for the lower rank case $D^{(2)}_2$ \cite{NePR19,RJS19,RPJS20,NeRe21a,RoJS21,FrGe22,FrGe23}. It might also be interesting to investigate the different parameter regimes $\gamma\in (\frac{\pi}{4},\frac{\pi}{2})$. Another natural but challenging topic might be the generalization to $D^{(2)}_n$ with $n>3$, as has been done for the $A^{(2)}_n$ series \cite{VeJS16a}. 
\section*{Acknowledgement}
  The authors thank Yacine Ikhlef, Gleb A. Kotousov and Marcio J. Martins for valuable discussions. HF and SG acknowledge funding provided by the Deutsche Forschungsgemeinschaft (DFG) under grant No.\ Fr~737/9-2 as part of the research unit   \textit{Correlations in Integrable Quantum Many-Body Systems} (FOR2316). RN was supported in part by
   the National Science Foundation under Grant No. NSF 2310594 and by a Cooper fellowship. ALR was supported by a UKRI Future Leaders Fellowship (grant number MR/T018909/1). Part of  the numerical work has been performed on the LUH compute cluster, which is funded by the Leibniz Universität Hannover, the Lower Saxony Ministry of Science and Culture and the DFG. 

\appendix

\section{Proofs of symmetries}\label{sec:proofs}

We sketch here proofs of some of the symmetries noted in the main text.

\subsection{Crossing symmetry of the transfer matrix \eqref{crossing}}

To prove that the transfer matrix \eqref{twistedtransf} has
the crossing symmetry \eqref{crossing}, we begin by noting that the \emph{transposed} transfer matrix can be expressed as
\begin{align}
    \mathbbm{t}^t(u; \{\phi_j \}) & = \mathrm{tr}_{0}\, \left( \mathbb{K}_0(\{\phi_j \})\, \mathbb{R}_{0L}(u)\dots \mathbb{R}_{01}(u) \right)^{t_0 t_1 \cdots t_L} \nonumber \\
    & = \mathrm{tr}_{0}\, \mathbb{R}_{01}^{t_0 t_1}(u) \cdots \mathbb{R}_{0L}^{t_0 t_L}(u)\, \mathbb{K}^{t_0}_0(\{\phi_j \}) \nonumber \\
    & = \mathrm{tr}_{0}\, \mathbb{K}_0(\{\phi_j \})\, \mathbb{R}_{10}(u)  \cdots \mathbb{R}_{L0}(u) \,,
    \label{lemma1}
\end{align}
where we have passed to the final line using the PT symmetry \eqref{PT} and the fact that the twist matrix \eqref{twistmatrix} is symmetric $\mathbb{K}^t = \mathbb{K}$.
We then observe that the transfer matrix itself can be expressed as 
\begin{align}
    \mathbbm{t}(u; \{\phi_j \}) & = \mathrm{tr}_{0}\, \left( \mathbb{K}_0(\{\phi_j \})\, \mathbb{R}_{0L}(u)\dots \mathbb{R}_{01}(u) \right)^{t_0} \nonumber \\
    & = \mathrm{tr}_{0}\, \mathbb{R}_{01}^{t_0}(u) \cdots \mathbb{R}_{0L}^{t_0}(u)\, \mathbb{K}^{t_0}_0(\{\phi_j \}) \nonumber \\
    & = \mathrm{tr}_{0}\, V_0^{t_0}\, \mathbb{R}_{10}(4\ri \gamma-u)\, V_0^{t_0} \cdots V_0^{t_0}\, \mathbb{R}_{L0}(4\ri \gamma-u)\, V_0^{t_0}\, \mathbb{K}_0(\{\phi_j \}) \nonumber \\
    & = \mathrm{tr}_{0}\, \mathbb{K}_0(\{-\phi_j \})\, \mathbb{R}_{10}(4\ri \gamma-u)\,  \cdots  \mathbb{R}_{L0}(4\ri \gamma-u)\,  \nonumber \\
    & = \mathbbm{t}^t(4\ri \gamma-u; \{-\phi_j \})\,.
    \label{crossingproof}
\end{align}
In passing to the third equality, we have used \eqref{crossingR} and \eqref{PT}; and in passing to the fourth equality, we have used the fact
\begin{equation}
    V\, \mathbb{K}(\{\phi_j \}) = \mathbb{K}(\{-\phi_j \})\, V \,.
    \label{VK}
\end{equation}
Finally, to pass to the last line of \eqref{crossingproof}, we have used the result \eqref{lemma1}. 

\subsection{\texorpdfstring{$W(0)$}{W(0)} symmetry of the transfer matrix \eqref{W_Sym}}

The transfer matrix \eqref{twistedtransf} transforms under $W(0)^{\otimes L}$ as
\begin{align}
   W(0)^{\otimes L}\,  
   \mathbbm{t}(u; \phi_1, \phi_2)\, W(0)^{\otimes L} & = W(0)^{\otimes L}\, \mathrm{tr}_{0}\, \left(  \mathbb{K}_0(\phi_1, \phi_2)\, \mathbb{R}_{0L}(u)\dots \mathbb{R}_{01}(u)\, \right) W(0)^{\otimes L} \nonumber \\
 & =  \mathrm{tr}_{0}\, \mathbb{K}_0(\phi_1, \phi_2)\, W_L(0)\, \mathbb{R}_{0L}(u)\, W_L(0)\, \dots W_1(0)\,\mathbb{R}_{01}(u)\, W_1(0) \nonumber \\
  & =  \mathrm{tr}_{0}\, \mathbb{K}_0(\phi_1, \phi_2)\, W_0(u)\, \mathbb{R}_{0L}(u)\, W_0(u)\, \dots W_0(u)\,\mathbb{R}_{01}(u)\, W_0(u) \nonumber \\
  & =  \mathrm{tr}_{0}\, \mathbb{K}_0(-\phi_2, -\phi_1)\,  \mathbb{R}_{0L}(u)\,  \dots \,\mathbb{R}_{01}(u) \nonumber \\
   & = \mathbbm{t}(u; -\phi_2, -\phi_1)\,,
   \label{Wsymproof}
\end{align}
where we have passed to the third equality using \eqref{RWsymmetry}, and to the fourth equality using the fact
\begin{equation}
    W(u)\, \mathbb{K}(\phi_1, \phi_2)\, W(u) =  \mathbb{K}(-\phi_2, -\phi_1) \,. 
\end{equation}

\subsection{CPT symmetry of the transfer matrix \eqref{CPT}}

In order to prove that the transfer matrix \eqref{twistedtransf} has
the CPT symmetry \eqref{CPT}, we begin by observing that the parity operator \eqref{parityop} acts as
\begin{align}
    \Pi\, \mathbbm{t}(u; \{\phi_j \})\, \Pi &= \mathrm{tr}_{0}\, \Pi \left(\mathbb{K}_0(\{\phi_j \})\, \mathbb{R}_{0L}(u)\dots \mathbb{R}_{01}(u) \right) \Pi \nonumber\\
    & = \mathrm{tr}_{0}\, \mathbb{K}_0(\{\phi_j \})\, \mathbb{R}_{01}(u)\dots \mathbb{R}_{0L}(u) \,.
\end{align}
It then follows that
\begin{align}
    V^{\otimes L}\, \Pi\, \mathbbm{t}(u; \{\phi_j \})\, \Pi\, V^{\otimes L} &= \mathrm{tr}_{0}\, \mathbb{K}_0(\{\phi_j \})\, V_1\, \mathbb{R}_{01}(u)\, V_1 \dots V_L\, \mathbb{R}_{0L}(u) \, V_L \nonumber \\
    &= \mathrm{tr}_{0}\,  \mathbb{K}_0(\{\phi_j \})\, V_0\, \mathbb{R}_{10}(u)\, V_0\, \dots  V_0\, \mathbb{R}_{L0}(u)\, V_0 \nonumber \\
    &= \mathrm{tr}_{0}\, V_0\, \mathbb{K}_0(\{\phi_j \})\, V_0\, \mathbb{R}_{10}(u)\,  \dots  \mathbb{R}_{L0}(u) \nonumber \\
    &= \mathrm{tr}_{0}\, \mathbb{K}_0(\{-\phi_j \})\, \mathbb{R}_{10}(u)\,  \dots  \mathbb{R}_{L0}(u) \nonumber \\
    &= \mathbbm{t}^t(u; \{-\phi_j \}) \,,
    \label{CPTproof}
\end{align}
where we have passed to the second equality using \eqref{Vprop}, to the fourth equality using \eqref{VK}, and to the last equality using \eqref{lemma1}. 

\subsection{CP symmetry of the Hamiltonian \eqref{CPHam}}

In order to prove that the Hamiltonian has the CP symmetry \eqref{CPHam}, we note the more explicit expression that follows from its definition \eqref{eq:Hamil} 
\begin{equation}
    \mathbb{H} \sim \mathbbm{t}^{-1}(0)\, \mathbbm{t}'(0) = \sum_{i=1}^{L-1} H_{i,i+1} +  \mathbb{K}^{-1}_L\, H_{L,1}\, \mathbb{K}_L \,, \qquad H_{i,i+1} = \mathcal{P}_{i,i+1}\, \mathbb{R}'_{i,i+1}(0) \,,
    \label{Hamexpl}
\end{equation}
and proceed to show that the ``bulk'' and ``boundary'' terms are separately invariant.

For the ``bulk'' terms in \eqref{Hamexpl}, we observe that parity acts as
\begin{equation}
    \Pi\, H_{i,i+1} \Pi = H_{L+1-i,L-i} = \mathcal{P}_{L+1-i,L-i}\, \mathbb{R}'_{L+1-i,L-i}(0) \,.
\end{equation}
It follows that CP acts as 
\begin{align}
  V^{\otimes L}\, \Pi\, H_{i,i+1} \Pi\, V^{\otimes L} &= \mathcal{P}_{L+1-i,L-i}\, V_{L-i}\, V_{L+1-i}\,\mathbb{R}'_{L+1-i,L-i}(0)\,  V_{L-i}\,  V_{L+1-i} \nonumber \\
  &= \mathcal{P}_{L+1-i,L-i}\,\mathbb{R}'_{L-i,L+1-i}(0)\,  \nonumber \\
  &= H_{L-i,L+1-i}\,,
\end{align}
where we have passed to the second equality using \eqref{Vprop}. Summing over $i$, we obtain
\begin{equation}
    V^{\otimes L}\, \Pi\, \left(\sum_{i=1}^{L-1} H_{i,i+1} \right)\, \Pi\, V^{\otimes L} = \sum_{i=1}^{L-1} H_{L-i,L+1-i} = \sum_{j=1}^{L-1} H_{j,j+1}\,,
\end{equation}
where we have performed the change of variables $j=L-i$
to pass to the final equality. In short, the ``bulk'' terms in \eqref{Hamexpl} are CP invariant.

For the ``boundary'' term in \eqref{Hamexpl}, we observe that parity acts as
\begin{equation}
    \Pi\, \mathbb{K}^{-1}_L\, H_{L,1}\, \mathbb{K}_L  \Pi = \mathbb{K}^{-1}_1\, H_{1,L}\, \mathbb{K}_1 \,,
\end{equation}
and therefor CP acts as
\begin{align}
    V^{\otimes L}\, \Pi\, \mathbb{K}^{-1}_L\, H_{L,1}\, \mathbb{K}_L  \Pi\, V^{\otimes L}  &= V_1\, V_L\, \mathbb{K}^{-1}_1\, H_{1,L}\, \mathbb{K}_1\, V_1\, V_L \nonumber \\
    &= \mathbb{K}_1\, V_1\, V_L\,  H_{1,L}\,  V_1\, V_L\, \mathbb{K}^{-1}_1 \nonumber \\
    &= \mathbb{K}_1\,  H_{L,1}\, \mathbb{K}^{-1}_1 \nonumber \\
    &= \mathbb{K}^{-1}_L\,  H_{L,1}\, \mathbb{K}_L \,,
\end{align}
where we have passed to the second equality using \eqref{VK}, to the third equality using \eqref{Vprop}, and to the final equality using the fact
\begin{equation}
    \left[ H_{L,1}\,, \mathbb{K}_L\, \mathbb{K}_1 \right] = 0 \,.
\end{equation}
We conclude that the ``boundary'' term is also CP invariant, and therefore, so is the full Hamiltonian \eqref{Hamexpl}.

\section{Introducing a diagonal twist in the Bethe ansatz}\label{sec:twistBA}

We present here a derivation of the
result \eqref{TQtwist} for the eigenvalue $\Lambda(u)$ of the (twisted) transfer matrix, starting from the untwisted result \cite{Resh87}
\begin{equation}
    \begin{aligned}
    \Lambda(u)\Big\vert_{\phi_1=\phi_2=0} =& \left(4\sinh(u-2\ri\gamma)\sinh(u-4\ri\gamma)\right)^{L} A(u)
               + \left(4 \sinh(u-4\ri\gamma) \sinh u\right)^L \sum_{\ell=1}^4 B_\ell(u)\,\\
               &+ \left(4\sinh(u-2\ri\gamma)\sinh u\right)^{L} C(u) \,,
    \end{aligned}
    \label{TQ}
\end{equation}
where $A(u), B_{\ell}(u), C(u)$ are given in Eqs. \eqref{AB1B2},\eqref{CB3B4}.

We proceed, in the same spirit as \cite{Resh87},
using a kind of analytical Bethe ansatz approach. The asymptotic behavior of the monodromy matrix \eqref{monodromy} is given by
\begin{equation}
     \mathbb{T}(u)\underset{u\rightarrow \infty}{\sim}  (\re^{2u-4\ri\gamma})^L \left\{ \text{diag}\left( \re^{-2\ri\gamma \mathbbm{h}_1}\,, \re^{-2\ri\gamma \mathbbm{h}_2} \,, \id \,, \id \,, \re^{2\ri\gamma \mathbbm{h}_2} \,, \re^{2\ri\gamma \mathbbm{h}_1}\right) + \ldots \right\}\,,
\end{equation}
where the ellipsis denotes off-diagonal terms that will not contribute to the final result.
It follows that the twisted transfer matrix \eqref{twistedtransf} has the asymptotic behavior 
\begin{equation}
     \mathbbm{t}(u)\underset{u\rightarrow \infty}{\sim}  (\re^{2u-4\ri\gamma})^L \left\{ 
    2\, \id + \sum_{j=1}^2 \left( \re^{\ri \phi_j} \re^{-2\ri\gamma \mathbbm{h}_j} + \re^{-\ri \phi_j} \re^{2\ri\gamma \mathbbm{h}_j} \right) + \ldots \right\} \,.
    \label{transfasym}
\end{equation}

We assume that the eigenvalue $\Lambda(u)$ is given by the following ``dressed'' version of the periodic result \eqref{TQ}
\begin{align}
    \Lambda(u) =& \left(4\sinh(u-2\ri\gamma)\sinh(u-4\ri\gamma)\right)^{L} a\, A(u)
               + \left(4 \sinh(u-4\ri\gamma) \sinh u\right)^L \sum_{\ell=1}^4 b_\ell\, B_\ell(u)\, \nonumber\\
               &+ \left(4\sinh(u-2\ri\gamma)\sinh u\right)^{L} c\, C(u) \,,
\label{TQtwist2}
\end{align}
where the parameters $a\,, b_\ell \,, c$ are still to be determined.
Evidently, we have from \eqref{simult1}
\begin{equation}
    \langle \Lambda | \mathbbm{t}(u) | \Lambda\rangle = \Lambda(u) \,.
    \label{expval}
\end{equation}
For $u\rightarrow\infty$, we evaluate the LHS of \eqref{expval} using \eqref{transfasym} and \eqref{simult2}, obtaining
\begin{align}
 \langle \Lambda | \mathbbm{t}(u) | \Lambda\rangle  
 \underset{u\rightarrow \infty}{\sim}   & (\re^{2u-4\ri\gamma})^L \Big\{ 
    \re^{\ri \phi_1} \re^{-2\ri\gamma (L-m_1)}
    +  \re^{\ri \phi_2} \re^{-2\ri\gamma (m_1-m_2)}\nonumber \\
   &  + 2
     + \re^{-\ri \phi_1} \re^{2\ri\gamma (L-m_1)}
    +  \re^{-\ri \phi_2} \re^{2\ri\gamma (m_1-m_2)}\Big\}
    \,;
    \label{LHS}
\end{align}
and we evaluate the RHS of \eqref{expval} using \eqref{TQtwist2} to obtain
\begin{align}
    \Lambda(u) \underset{u\rightarrow \infty}{\sim}   & 
    (\re^{2u-4\ri\gamma})^L \Big\{ 
    a\, \re^{-2\ri\gamma (L-m_1)}
    +  b_1\, \re^{-2\ri\gamma (m_1-m_2)}\nonumber \\
   &  + b_2 + b_3  
    +  b_4\, \re^{2\ri\gamma (m_1-m_2)}
    + c\, \re^{2\ri\gamma (L-m_1)}
 \Big\} \,.
 \label{RHS}
\end{align}
Comparing \eqref{LHS} and \eqref{RHS}, we conclude that the parameters are given by
\begin{equation}
    a = \re^{\ri \phi_1}\,, \quad b_1 = \re^{\ri \phi_2}\,, \quad b_2 = b_3 =1  \,, \quad b_4 = \re^{-\ri \phi_2}\,, \quad c = \re^{-\ri \phi_1}\,,
\end{equation}
which is the result in \eqref{TQtwist}.

	\providecommand{\href}[2]{#2}\begingroup\raggedright\endgroup
\end{document}